\documentclass[manuscript]{emulateapj}
\usepackage{natbib,epsfig,graphicx}

\slugcomment{Accepted for publication in Astronomical Journal}

\shorttitle{PN sub-populations in NGC 4697}

\newcommand{\mpc}{{\rm\,Mpc}}
\newcommand{\etal}{et al.}
\newcommand{\msun}{{\rm\,M_\sun}}
\newcommand{\kms}{{\rm\,km\,s^{-1}}}
\def\spose#1{\hbox to 0pt{#1\hss}}
\def\lta{\mathrel{\spose{\lower 3pt\hbox{$\mathchar"218$}}
     \raise 2.0pt\hbox{$\mathchar"13C$}}}
\def\gta{\mathrel{\spose{\lower 3pt\hbox{$\mathchar"218$}}
     \raise 2.0pt\hbox{$\mathchar"13E$}}}

\begin{document}

\title{Kinematic evidence for different Planetary Nebulae Populations\\ 
in the elliptical galaxy NGC 4697}

\author{Niranjan Sambhus} 
\affil{Astronomisches Institut, Universit\"{a}t
Basel, Venusstrasse 7, 4012 Binningen, Switzerland}
\email{sambhus@astro.unibas.ch}

\author{Ortwin Gerhard}
\affil{Current address: Max-Planck Institut f\"{u}r Extraterrestrische
Physik, Giessenbachstrasse, D-85748 Garching, Germany}
\affil{Astronomisches Institut, Universit\"{a}t
Basel, Venusstrasse 7, 4012 Binningen, Switzerland}
\email{gerhard@mpe.mpg.de}

\author{Roberto H. M\'endez}
\affil{Institute for Astronomy, University of Hawaii, 2680 Woodlawn Drive, 
Honolulu, HI 96822, USA}
\email{mendez@ifa.hawaii.edu}

\begin{abstract}
We have analysed the magnitudes, kinematics and positions of a
complete sample of 320 PNs in the elliptical galaxy NGC 4697. We show
(i) that the PNs in NGC 4697 do not constitute a single population
that is a fair tracer of the distribution of all stars. The radial
velocity distributions, mean velocities, and dispersions of bright and
faint subsamples differ with high statistical confidence. (ii) Using
the combined data for PNs brighter than $26.2$, we have identified a
subpopulation of PNs which is azimuthally unmixed and kinematically
peculiar, and which thus neither traces the distribution of all stars
nor can it be in dynamical equilibrium with the galaxy potential.
(iii) The planetary nebula luminosity functions (PNLF) of two
kinematic subsamples in NGC 4697 differ with 99.7\% confidence, ruling
out a universal PNLF. We estimate that the inferred secondary PN
population introduces an uncertainty in the bright cutoff magnitude of
$\sim0.15$ mag for this galaxy.  -- We argue that this secondary PN
distribution may be associated with a younger, $\gta 1$ Gyr old
stellar population, perhaps formed in tidal structures that have now
fallen back onto the galaxy, as has previously been suggested for the
X-ray point sources in this galaxy, or coming from a more recent
merger/accretion with a red galaxy. The use of PNs for extragalactic
distance determinations is not necessarily compromised, but their use
as dynamical tracers of dark halos will require deep observations and
careful analysis of large PN samples.
\end{abstract}


\section{Introduction}

Planetary Nebulae (PNs) have become increasingly important in
extragalactic astronomy, for distance determinations via their
luminosity function (LF) \citep[and references therein]{jac89,cia89,
jcf90,men99,fer00,cia03}, as kinematic tracers of the dark halos of
galaxies \citep{arn98,sag00,men01,rom03}, and as tracers for the
distribution and kinematics of the diffuse stellar population in
galaxy clusters \citep{fel98,fel04,arn02,arn04,ger05}.  Due to their
strong narrow line emission at $[\mathrm{OIII}] \lambda 5007$, PNs can
be easily detected out to distances beyond $20 \mpc$ with narrow-band
photometry and slitless spectroscopy \citep{fel98,arn02,men01,dou02},
and to $\sim 100 \mpc$ with multi-slit imaging spectroscopy
\citep{ger05}. Moreover, they are observed in elliptical and spiral
galaxies, making them an indispensible tool to support distances
obtained by other methods (such as Cepheids, Surface brightness
fluctuations, the Tully-Fisher relation, SNe Ia), and to measure the
kinematics of stellar populations whose surface brightness is too
faint for absorption line spectroscopy.

For distance determination the Planetary Nebulae Luminosity Function
(PNLF) is normally modeled as having a universal shape that depends
only on the absolute bright magnitude cutoff $M^{\star}$:
\begin{equation}
N(M) \propto e^{0.307 M} (1 - e^{3(M^{\star} - M)}) \,,
\label{pnlfeqn}
\end{equation}

\noindent where $N(M)$ is the number of PNs with absolute magnitude $M$
\citep{cia89}. Observationally, the cutoff magnitude $M^{\star}$ has a 
quasi-universal value of $-4.5$ with only a weak dependence on host galaxy
metallicity expressed by the system's oxygen abundance, and which can be 
compensated for by a quadratic relation in $[\rm{O/H}]$ \citep{dop92,cia02}. 
In practice, the PN magnitudes $m(5007)$, after correcting for the 
interstellar reddening, are fitted to the model PNLF of eq.~\ref{pnlfeqn}
convolved with the photometric error profile, yielding a value of the 
distance modulus \citep{cia89}. The absence of any systematic variations 
in $M^{\star}$ and the PNLF shape has been verified in galaxies with 
significant population gradients, and among galaxies of different 
morphologies within galaxy clusters/groups up to Virgo 
\citep[and references therein]{jac97,cia03}.

This universality of the PNLF and the cutoff magnitude $M^{\star}$ must
be considered surprising, given that the PN luminosity in the
$[\mathrm{OIII}] \lambda 5007$ line depends on the mass and
metallicity of the central star, as well as on the electron gas
temperature, optical thickness and dust extinction of the surrounding
nebula. Indeed, some current semi-analytic simulations of the PNLF
seem to be at odds with the observational trends. \citet{men93,men97}
indicate small possible dependencies of $M^{\star}$ on the total size
of the PN population, on the time elapsed since the last episode of
star formation, and on how optically thin the PNs are; concluding,
however, that only careful studies would detect such effects in the
observed PNLF.  In contrast, more recent PNLF simulations by
\citet{mar04} contradict the observed narrow spread in $M^{\star}$ and
predict large variations of several magnitudes depending on a variety
of realistic star formation and evolution scenarios. So is the PNLF
truly quasi-universal and its cutoff magnitude nearly independent of 
population age and metallicity? 
 
PNs are also important as test particles to study the kinematics and
dark matter distribution in the halos of elliptical galaxies. Since
the PN population is expected to arise from the underlying galactic
stellar distribution, their radial velocities can be used as effective
kinematic tracers of the mass distribution. However, the required PN
sample sizes are many 100s \citep{ms93}, or at least 100 or more in
conjunction with absorption line spectroscopy, which has limited this
application to only a few nearby galaxies
\citep{hui95,arn98,sag00,men01,rom03,pen04}. In recent simulations
of disk galaxy mergers involving dark matter, stars, and gas,
\citet{dek05} predict that the young stars formed in the
merger have steeper density profiles and larger radial anisotropy than
the old stars from the progenitor galaxies, and they argue that if the
PNs observed in elliptical galaxies were to correspond to the young
population rather than to all stars in the simulations, their velocity
dispersion profile would match the measured dispersion profiles of
\citet{rom03}. So do PNs really trace the stars and their kinematics
in elliptical galaxies?

Different stellar populations may have, and in general would have,
different phase-space distributions in the same galaxy potential. The
simplest approach for dynamical modelling, taking the PN velocities as
a random sampling of the stellar velocities, is however valid only
when the PN population properties and their kinematics are
uncorrelated. Except in special cases this also requires that the PNLF
is independent of the stellar population. Vice-versa, if there existed
differences in the PNLF or the bright cutoff magnitude for different
stellar populations, they would best be identified by studying the
correlations between PN magnitudes and kinematics or positions of
these tracers, in a single galaxy where all PNs are at the same
distance.

In this paper, we report on such a study in the elliptical galaxy NGC
4697, an excellent target for this purpose because of the large sample
of PN velocities known from \citet{men01}.  Our analysis shows the
existence of distinct PN populations which differ in their kinematics,
brightnesses, and spatial distributions. This suggests that the answer
to both the questions posed above may be 'no' -- in general, different
stellar populations may have slightly different PNLFs, and the
observed PN population in elliptical galaxies may not be a fair tracer
of their stars. The paper is organised as follows: in \S~\ref{data} we
review the properties and PN data of this galaxy and discuss the
magnitude and velocity completeness of our sample. Our statistical
analysis of these data is given in \S~\ref{analysis} where we
demonstrate the inhomogeneity of the sample in the space of
velocities, magnitudes, and positions. \S\S~\ref{discn},\ref{result}
conclude our work, giving also a brief discussion of its implications.

\section{NGC 4697: properties and PN sample}
\label{data}

NGC 4697 is a normal, almost edge-on E4-5 galaxy located along the
Virgo southern extension. From the multi-colour CCD photometry of
\citet{gou94}, the effective radius is $\rm{R_{eff}} = 95\arcsec$, the
mean ellipticity is 0.45, and the PA is constant, consistent with a
near-axisymmetric luminosity distribution.  Isophotal analysis shows
that this galaxy has a positive $a4$ coefficient suggesting a
disk-like structure within $0.6 \rm{R_{eff}}$
\citep{car87,sco95}.

From optical spectroscopy, its dominant stellar population has an age
of $\sim 9$ Gyr \citep{tra00}, consistent with the red mean B-V=0.91
colour.  Stellar absorption line kinematics along the major axis (PA
$66\degr$) of NGC 4697 have been reported by \citet{bin90} and
\citet{dej96}; these velocity data can be well described by dynamical
models based on the luminous mass distribution only.

\citet{men01} detected and measured radial velocities of 531 PNs extending
out to $300\arcsec (\sim 3 \rm{R_{eff}})$ in this galaxy, with
observational errors of $\sim 40 \kms$. Via dynamical analysis, they
determined a constant mass to light ratio $\rm{\Upsilon_B} = 11$
within $\sim 300\arcsec$, which is consistent with a $10$ Gyr old
stellar population with a Salpeter mass function and slightly super
solar metallicity.

X-ray observations with ROSAT \citep{san00} show only small amounts of hot
gas in the halo of this galaxy. Using more recent \emph{Chandra} data,
\citet{sar00} could resolve most of the X-ray emission into nonuniformly
distributed X-ray binary point sources (XPS), suggesting that NGC 4697
has lost most of its interstellar gas. Though NGC 4697 does not show
any signature of recent interactions, \citet{zez03} present evidence
that the distribution of these X-ray sources is inconsistent with the
optical morphology of NGC 4697, and propose that these sources are
mostly high mass X-ray binaries (HMXBs) associated with young stellar
populations related to fallback of material in tidal tails onto a
relaxed merger remnant, or to shock-induced star formation along tidal
tails. They estimate typical fallback times of such tidal tails to be
much longer than the settling timescale of the remnant and expect
similar results for other elliptical galaxies with populations of
$\sim 10$ Gyr age.

For the work in this paper we use the PN sample presented by
\citet{men01}. After removing the possible contaminants and unclear
detections, they report unambigous detection of 535 PNs. However only
526 out of 535 PNs have confirmed measurements of velocity \emph{and}
magnitude, and we use these in our analysis. In order to determine the
PNLF, it is crucial to estimate the magnitude where PN detection
incompleteness sets in. Detectability of a PN varies with the
background galaxy surface brightness; for a statistically complete
sample the surface number density of PNs should be directly
proportional to it. \citet{men01} show that their PN sample is
statistically complete down to $m(5007) = 27.6$ magnitudes outside an
elliptical region of semi-major axis $60\arcsec$. In our analysis, we
have thus defined two data sets: a {\sl Complete sample} (with PNs
brighter than $27.6$, outside the central ellipse of semi-major axis
$60\arcsec$), and a {\sl Total sample} (consisting of all PNs with
measured magnitude and radial velocity). The total number of PNs in
these data sets is 320 and 526, respectively.

The systemic velocity ($\rm{V_{sys}}$) of NGC 4697, obtained by
averaging the observed velocity of all 526 PNs is $1274 \kms$, which
agrees with the values quoted in the literature \citep[and references
therein]{men01}. The on-band filter configuration used to detect and
measure velocities of these PNs has a peak wavelength of $5028$ \AA,
peak transmission of $0.76$, equivalent width of $48.5$ \AA, and FWHM
of $60$ \AA\, \citep{men01}. The FWHM corresponds to a velocity range
of $\pm 1800 \kms$ around the systemic velocity of NGC 4697. Thus the
filter transmission width is large enough to facilitate observations
of PNs with all velocities bound to NGC 4697, irrespective of their
magnitude. Indeed, even at magnitudes as faint as $m(5007)=28$ in the
Total sample, PNs with large velocities $\sim 300 \kms$ are
detected. Thus the velocity coverage in both samples (Total and
Complete) is not biased with respect to the PN magnitudes.

The PN magnitudes were measured by \citet{men01} from their
undispersed images; they are accurate to 0.1 and 0.2 mag for $m(5007)$
brighter and fainter than 26.5, with systematic effects below 2\%.
As a further test relevant for the present work, \citet{men01}
used the redundancy provided between their E and W fields: plotting
magnitude differences between the two measurements (E and W) of PN
candidates as a function of difference in distance from the center
of the CCD, they found a scatter diagram without any evidence of
correlation. \citet{men01} estimated the errors in the PN velocities from
calibration, image registration, spectrograph deformation and guiding
errors to be $40\kms$. The velocities of 165 PNs were measured
independently in the E and W field of \citet{men01}; these velocities
agree within a standard deviation of $36\kms$. In order to check
whether a systematic difference between the velocities of bright and
faint PNs could be introduced by an asymmetric PSF (a possibility
suggested by K.~Freeman), we have superposed the PSF's of three groups
of 10 of the brightest PNs, one selected at random, and two selected
among those PNs with the highest and lowest radial velocities.  In the
three cases we estimated the shift of the centroid of the entire PSF
with respect to the centroid of the upper part. The shifts were
smaller than $10\kms$, and in some cases they were in the opposite
sense compared to the results discussed below.

\section{The distribution of NGC 4697 PNs in velocity, magnitude,
and position}
\label{analysis}

In this Section, we search for stellar population effects in the
kinematics of the PNs in NGC 4697, by analysing the Total and Complete
data sets with respect to their three observables: velocity, magnitude
and position.  For both data sets, we convert the observed PN radial
velocities into co-rotating or counter-rotating velocities, as
follows. With the galaxy center at the origin of the reference frame,
and the X-axis oriented along the major axis (PA$=66$ deg), the
absorption line stellar-kinematic data predict positive line-of-sight
mean velocity with respect to the galaxy systemic velocity, at slit
positions towards the South-West of the center with X coordinate $>
0$, and vice versa. We denote this sense of rotation as
\emph{co-rotating}, and the opposite sense as
\emph{counter-rotating}. By definition, the major axis absorption line
data is \emph{co-rotating}.

\subsection{Sub-populations of PNs in the velocity--magnitude plane}
\label{subpop}

\begin{figure*}
\centerline{
\includegraphics[width=0.593\linewidth]{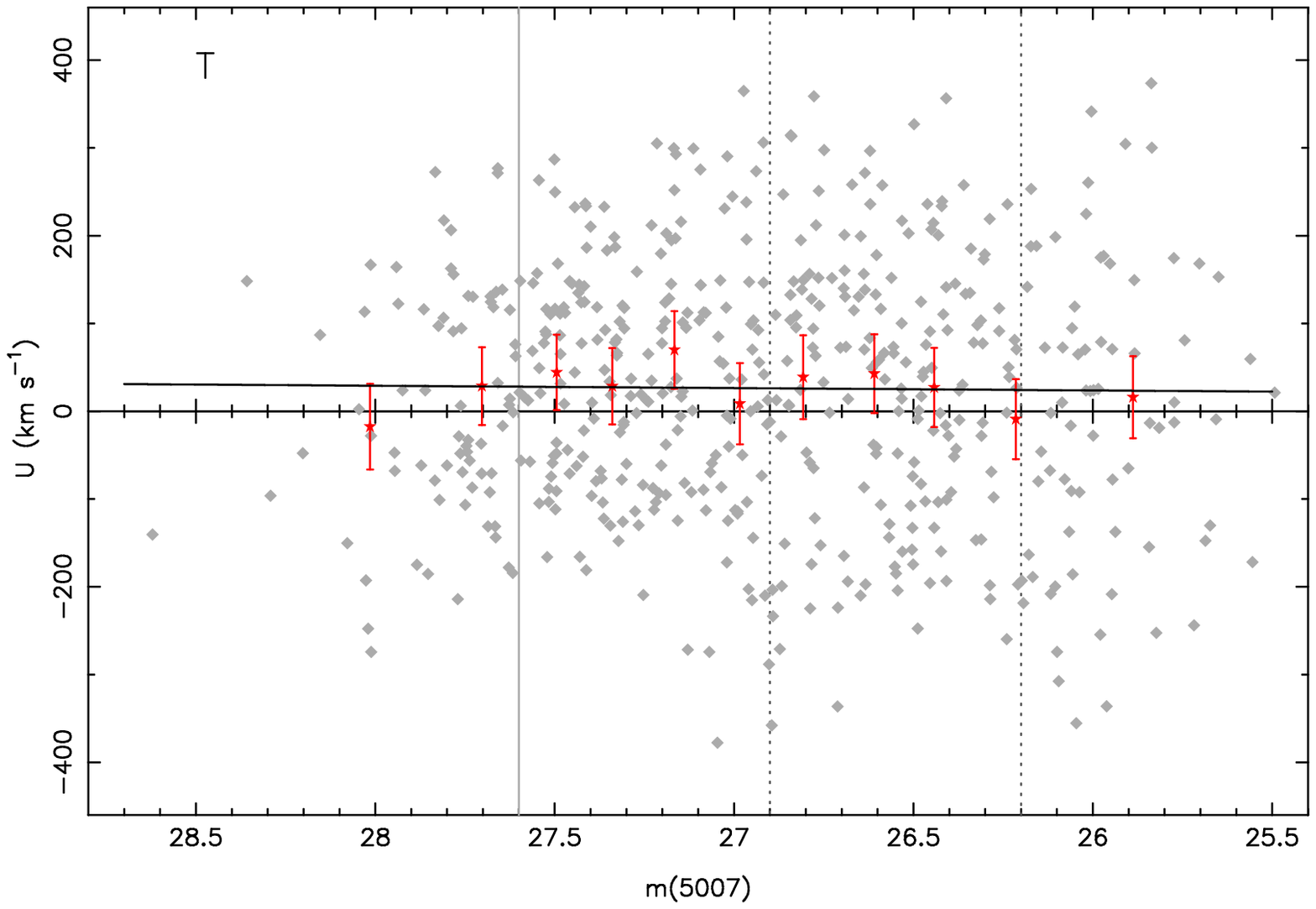}
\includegraphics[width=0.407\linewidth]{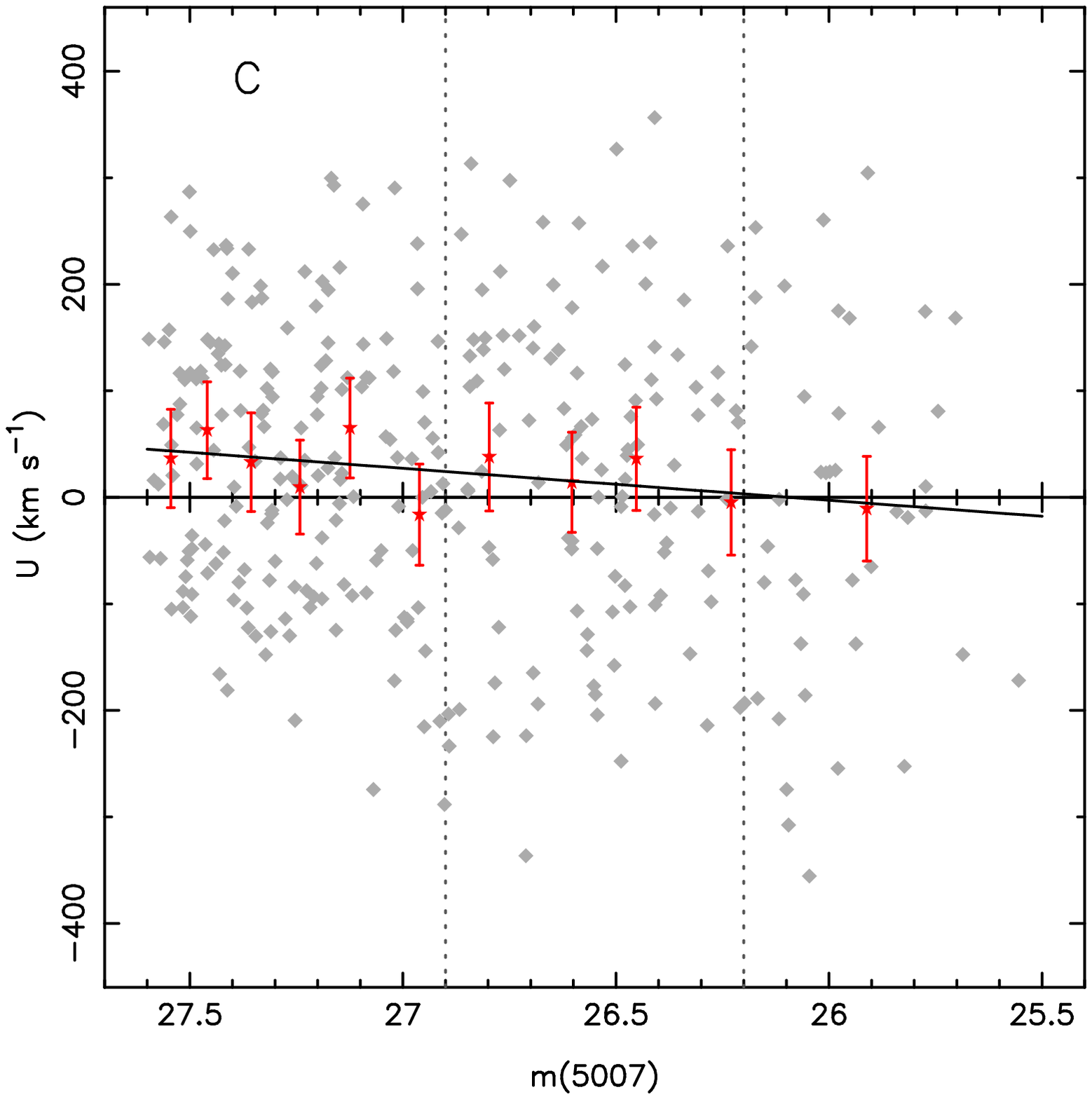}}
\caption{PN reduced velocity $U$ versus magnitude plots for the Total (T) 
and Complete (C) samples. Reduced velocities have been \emph{folded}
such that positive (negative) velocity denotes co- (counter-) rotating
PNs. Red stars (with error bars) are mean $U$ values of PNs binned in
magnitude such that each bin contains $30$ ($50$) data points in the
Complete (Total) sample. The error bars include the sample variance in
the bin and the velocity errors ($40 \kms$) added in quadrature. The
black solid lines are least square linear fits to this binned data. In
both samples, PNs with $m(5007) \la 26.5$ have mean $U \approx 0$. The
dotted vertical lines separate the Complete sample into 3 bins, each
of size 0.7 magnitude (see text). The Complete data set shows a linear
correlation between $m(5007)$ and $U$: Fainter PNs show more
co-rotation than their brighter counterparts. In the Total sample, the
vertical solid line at $m(5007) = 27.6$ denotes the limiting
magnitude, though not all PNs brighter than this magnitude constitute
the Complete sample.}
\label{mvplot}
\end{figure*}

After subtracting the systemic velocity from the PN radial velocities,
we define reduced velocities $\rm{U = (V - V_{sys}) * sign(x)}$ and
denote the PNs with $\rm{U}$ $>$ ($<$) 0 as co-rotating
(counter-rotating). By definition, the major axis absorption line
data have $\rm{U}>0$. The resulting values of $\rm{U}$ are displayed
against the observed magnitudes in Figure~\ref{mvplot}. Even at
magnitudes as faint as $m(5007)=28$ in the Total sample, PNs with
large velocities $\sim 300 \kms$ are detected, showing that there is
no kinematic bias at faint magnitudes. Henceforth, unless stated
otherwise, we will always use the Complete sample for our analysis.

\begin{figure}
\includegraphics[width=0.87\linewidth]{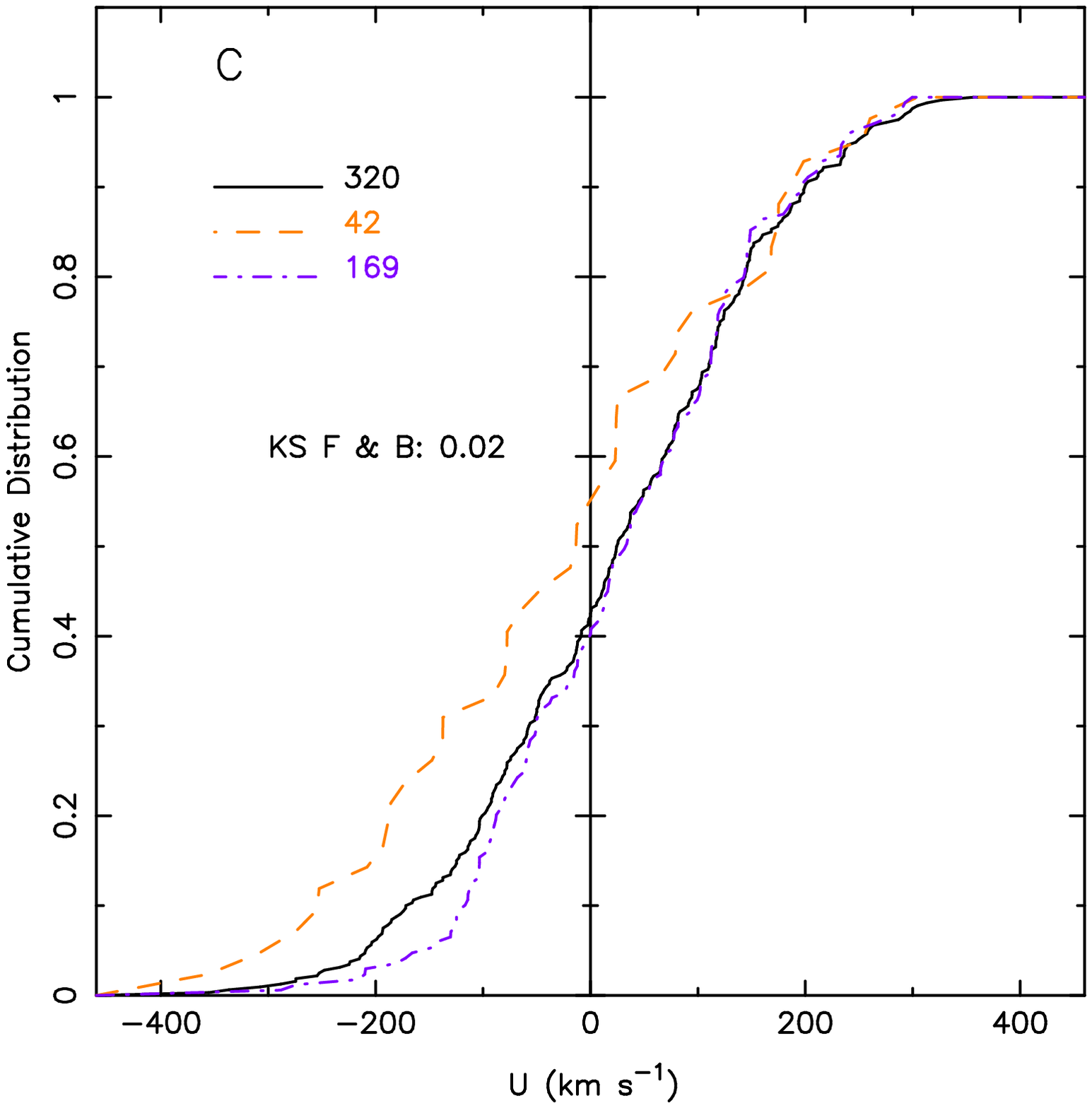}
\includegraphics[width=0.88\linewidth]{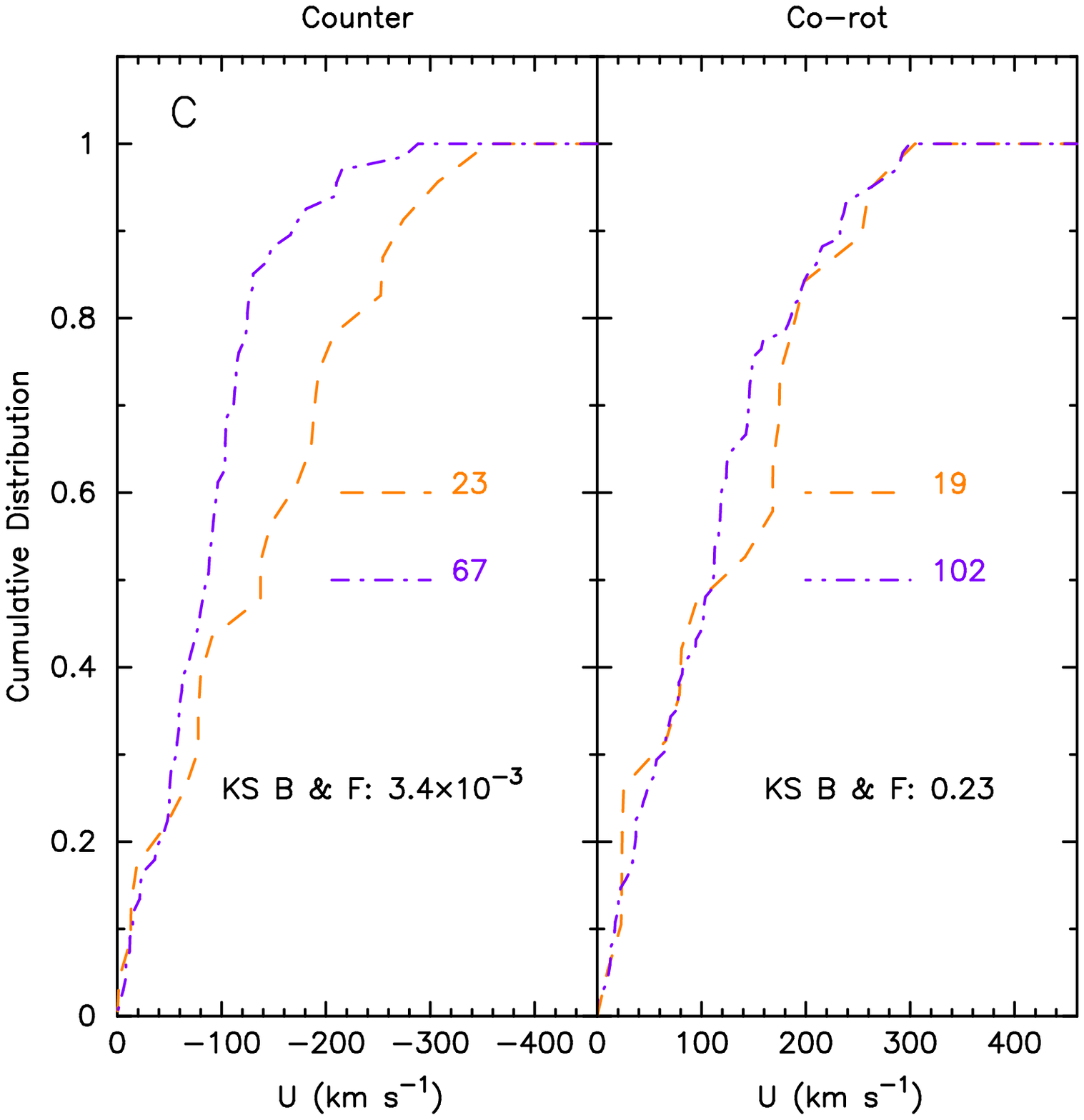}
\includegraphics[width=0.87\linewidth]{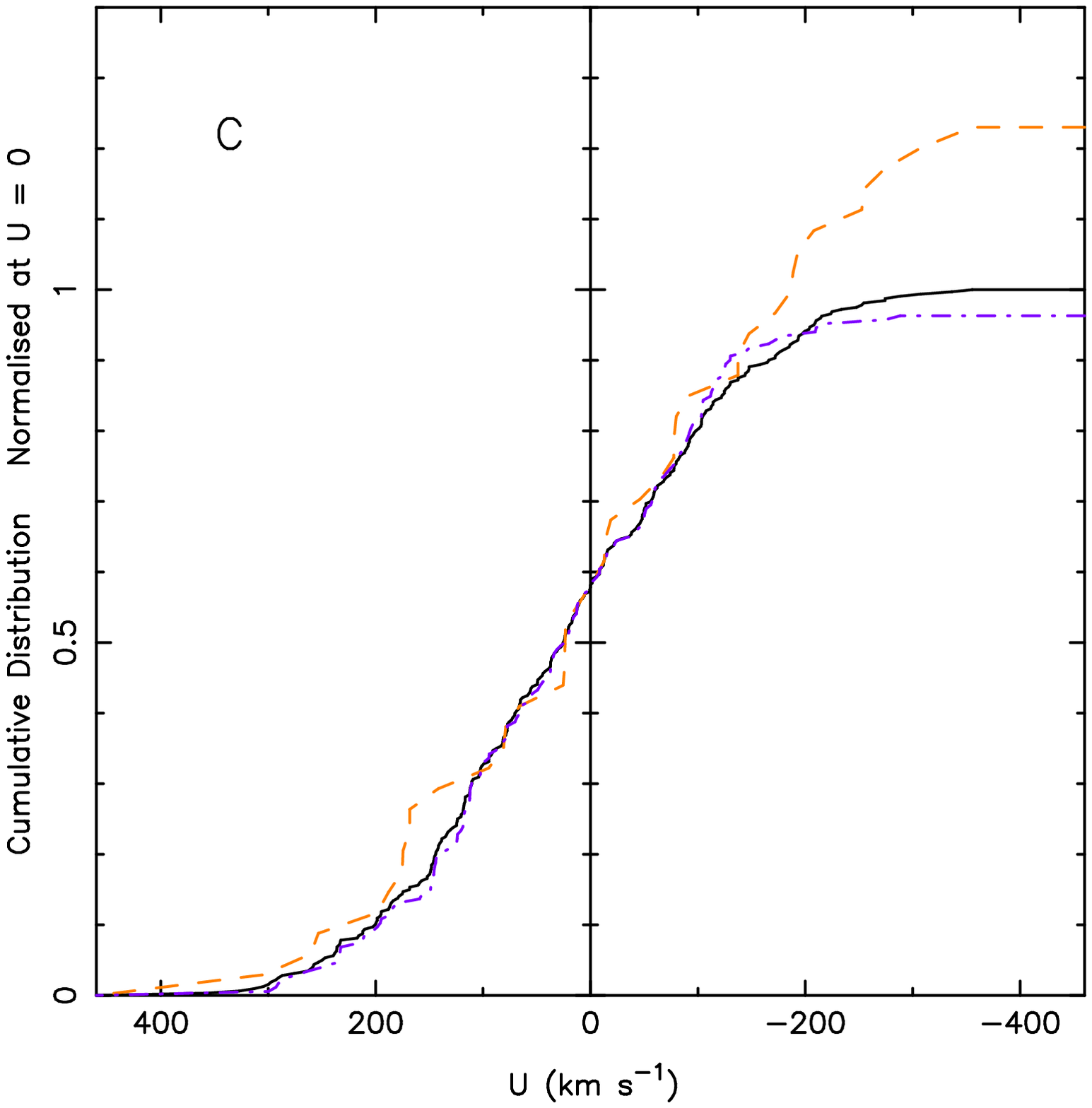}
\caption{Cumulative velocity distribution of brightest PNs ($m \leq 26.2$, 
orange dashed line), faintest PNs ($m \geq 26.9$, purple dashed-dotted
line), and the entire Complete sample (black solid line). The entire
velocity range is shown in the top panel. In the middle panel the
velocities are divided according to their sense of rotation. The
bottom panel shows the cumulative distribution (shown from $+ve$ to
$-ve$ values), normalised at $U = 0$. The Kolmogorov Smirnov
probabilities show that the brightest counter rotating PNs have
significantly different velocity distribution from the rest of the
PNs.}
\label{6bins}
\end{figure}

Figure~\ref{mvplot} shows that the Complete PN sample appears to
exhibit a correlation between magnitudes and kinematics, with faint
PNs showing more co-rotation than bright PNs. We have performed
several statistical tests to verify the significance and look for the
origin of this correlation.  Table~\ref{tabpear} shows the results of
Pearson's r-test for correlated data. Velocities of counter-rotating
PNs are strongly linearly correlated with their brightness, while
those of co-rotating PNs are almost independent of their magnitude
distribution. Further, we divided our sample into 3 equal magnitude
bins each of size $\Delta m = 0.7$\,, hereafter referred to as {\sl
faintest}, {\sl intermediate}, and {\sl brightest PNs}, and computed
the mean reduced velocity and its dispersion in each of these bins
along with the significance of their differences. As shown in
Table~\ref{tabvtftest}, the {\sl faint} and {\sl bright PN subsamples}
defined through these bins have different mean reduced velocity and
dispersions at $94\%$ and $99.3\%$ confidence, respectively.  In
Figure ~\ref{6bins} we have plotted the cumulative velocity
distribution of PNs in the brightest and faintest magnitude
bins. There is a visible excess of bright PNs with counter rotating
velocities. It is particularly evident from this figure that the
brightest counter rotating PNs display a velocity distribution that
differs from the rest of the PNs from the Complete sample with high
confidence.

Thus it is clear that the observed correlation between the PNs kinematics 
and their magnitudes is compelling, and it arises because the faintest and 
brightest PNs have significantly different velocity distributions. There 
appears to exist an additional component of bright, counter-rotating PNs 
with respect to the overall sample. 

\begin{table}[t]
\begin{center}
\begin{tabular}{|c|c|c|}
\hline
 & r & $\mathcal{P}_{\mathrm{r}}$ \\
\hline
All & $0.12$ & $3.1 \times 10^{-2}$ \\
Co-rot & $-3.5 \times 10^{-2}$ & $0.64$ \\
Counter-rot & $0.24$ & $5.8 \times 10^{-3}$ \\
\hline
\end{tabular}
\end{center}
\caption{Pearson's r-test for linear correlation of PN magnitudes
with reduced velocity ${\rm U}$, for the entire Complete sample,
the co-rotating, and the counter-rotating subsamples. 
Values of r close to $\pm 1$ indicate a strong linear correlation; values 
close to $0$ indicate little or no correlation. $\mathcal{P}_{\mathrm{r}}$ 
is the probability that two uncorrelated variables would give the 
r-coefficient as large as or larger than the measured one, for a normal 
distribution of r. Small values of $\mathcal{P}_{\mathrm{r}}$ imply 
significant correlation.}
\label{tabpear}
\end{table}

\begin{table}[b]
\begin{center}
\begin{tabular}{|c|c|c|c|c|c|c|}
\hline
 & $\overline{U}$ (error)& $\sigma$ & t &
$\mathcal{P}_t$ & F & $\mathcal{P}_F$\\
\hline
$m(5007) \geq 26.9$  & $32.9\, (9.5)$ & $123.1$ & & & & \\
 & & & $1.92$ & $6.0 \times 10^{-2}$ & $1.84$ & $7.4 \times 10^{-3}$  \\
$m(5007) \leq 26.2$ & $-20.0\, (25.8)$ & $167.3$ & & & & \\
\hline
\end{tabular}
\end{center}
\caption{Mean reduced velocity ($\overline{U}$) and dispersion 
($\sigma$) of brightest and faintest PNs, along with the results
from Student t-test and F-test. Very small values of $\mathcal{P}_t$
and $\mathcal{P}_F$ imply that the differences between the observed
$\overline{U}$ and $\sigma$ are statistically significant.}
\label{tabvtftest}
\end{table}

\subsection{Spatial Distribution}

If these correlations have a physical origin, they should also be
manifest in the spatial distribution of these PNs. Thus we now enquire
whether the PN kinematics and magnitudes depend on their spatial
location in the galaxy. 

\begin{figure}
\plotone{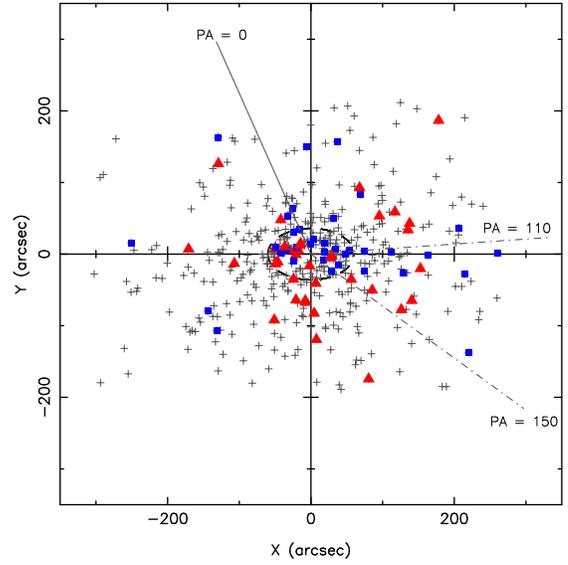}
\caption{Spatial distribution of 526 PNs in NGC 4697 (plus-symbols). PNs
brighter than $26.2$ are shown as filled blue squares (co-rotating)
and filled red triangles (counter-rotating). The dashed ellipse
denotes the central incompleteness region. The left-right asymmetry in
the distribution of bright PNs is apparent. The true North direction
is shown as a solid line along $\mathrm{PA}= 0\degr$. Following
\citet{zez03}, we measure the azimuth from this direction in the
clock-wise sense. Two special directions $\mathrm{PA} = 110\degr$ and
$150\degr$ are shown as dash-dotted lines (see text).}
\label{tspace}
\end{figure}

In Figure~\ref{tspace} we plot the spatial locations of all the 526
PNs in this galaxy. The central incompleteness ellipse is also
displayed. PNs brighter than $m = 26.2$ (which is $0.71$ magnitude
deeper than the brightest PN) are shown as filled blue squares
(co-rotating) and filled red triangles (counter-rotating). Inside the
incompleteness ellipse, the distribution of the bright PNs appears to
be concentrated around an elliptical annulus. However, we did not find
any kinematic evidence (like a rotation curve signature) relating
these bright PNs to the central stellar disk: either they are not
physically related to the disk, or the evidence from the data is
inconclusive. 

Outside the incompleteness ellipse, the distribution of bright PNs
does not follow the surface brightness of the host galaxy: there is a
significant left-right asymmetry, with more bright PNs to the right
side of the galaxy minor-axis ($27\pm 5$) than to the left side
($15\pm4$). \citet{men01} discuss at length the possible differences
in their E (east) and W (west) fields, and are convinced that the
maximum systematic errors in the measured photometry, positions and
radial velocities are below $2\%$, $0\arcsec.06$ and $20
\kms$, respectively. Hence we conclude that the left-right asymmetry
in number counts of bright PNs is not affected by detection
uncertainties. 

Subsequently, we carried out several tests to check whether the
brightest and faintest, or the co- and counter-rotating PNs are
distributed differently in the galaxy. It turns out that the radial PN
distribution is independent of their sense of rotation. However, the
distribution of PN distances from the galaxy mid-plane differs for co-
and counter-rotating PNs at a confidence level of $73\%$.

\begin{figure}
\plotone{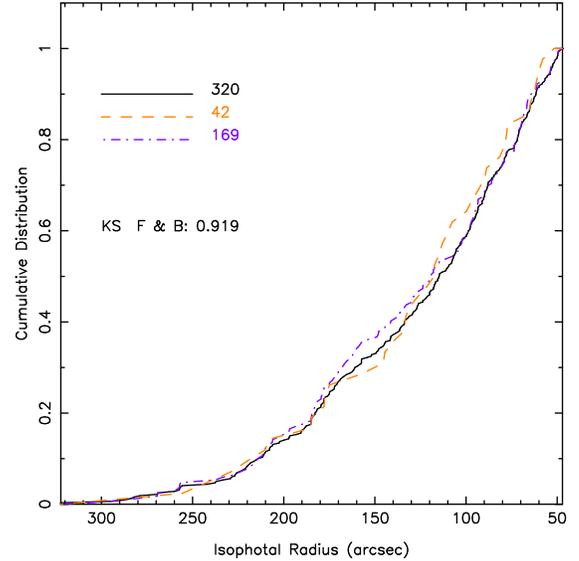}
\epsscale{1}
\caption{Cumulative radial distributions of all PNs in the
Complete sample (solid black line), the bright subsample (dashed
orange line), and the faint subsample (purple dash-dotted line).
The KS probability shows that these are all consistent with 
one another.}
\label{radial}
\end{figure}

\begin{figure}
\includegraphics[width=0.95\linewidth]{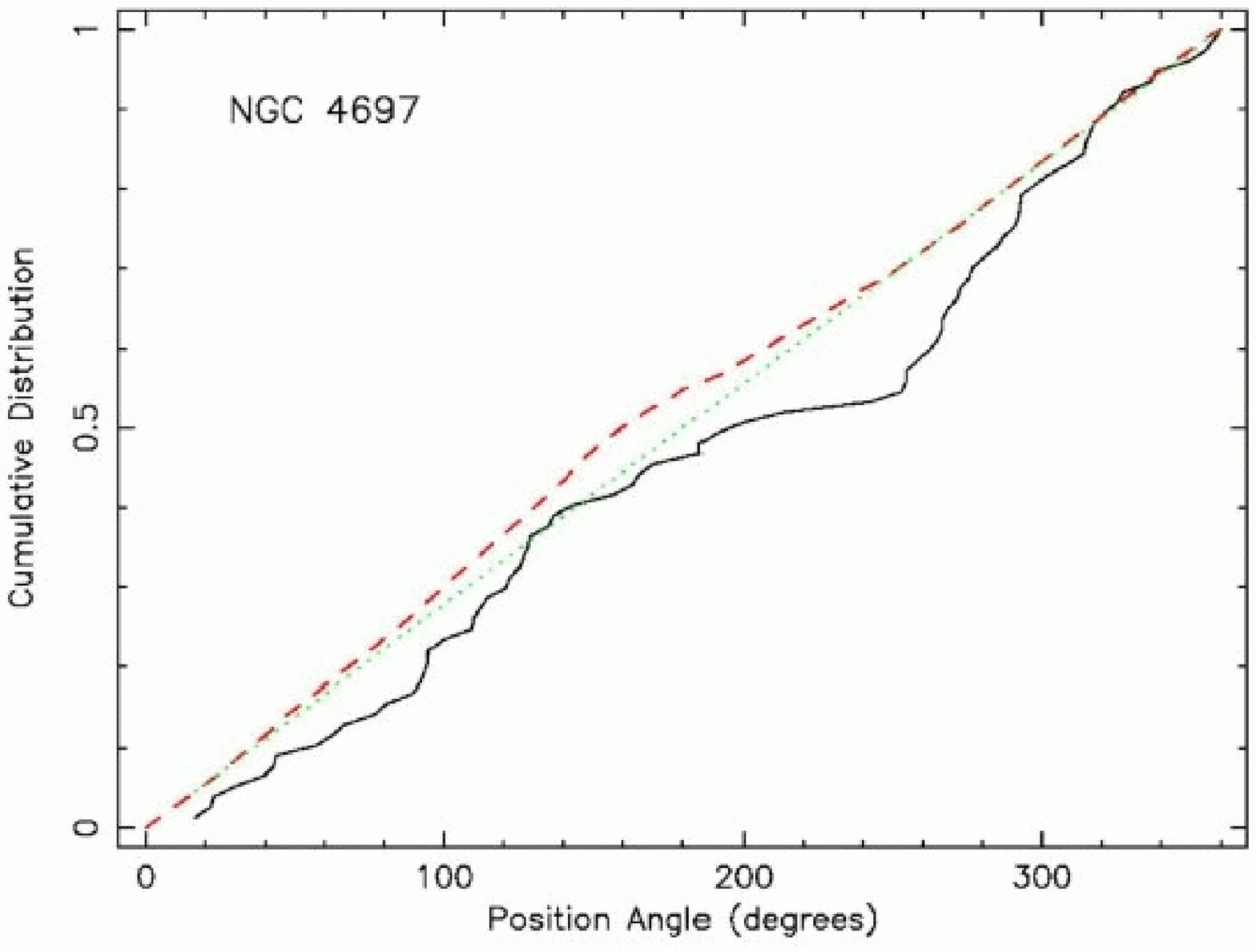}
\includegraphics[width=0.92\linewidth]{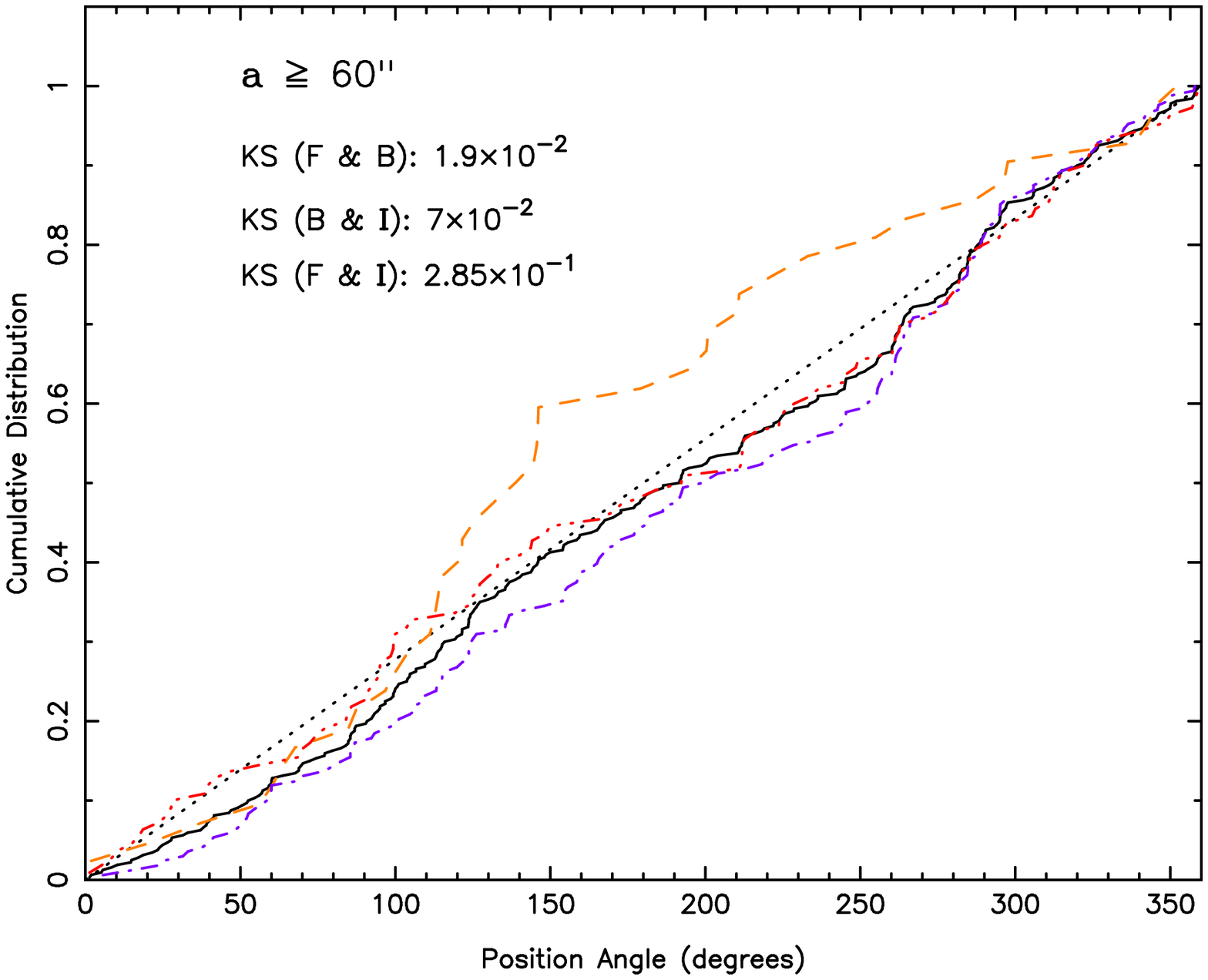}
\caption{Azimuthal distribution of discrete sources in NGC 4697. 
\emph{Top:} Cumulative distribution of azimuths of X-ray point
sources (black solid line), adapted from \citet{zez03}. The green
dotted line is for a uniform distribution, while the red dashed line
shows the azimuthal distribution of the galaxy optical light (in
R-band). \emph{Bottom:} PNs in the brightest, intermediate, and
faintest bins as defined in \S~\ref{subpop} are shown by the orange
dashed line, the red dash-triple dotted line, and the purple dash-dotted
line, respectively. All PNs in the Complete sample are shown by the
solid black line. Note the large increase in the distribution of
bright PNs and the moderate increase in the distribution of
intermediate PNs, relative to the faint PNs, at $\mathrm{PA} \sim
110\degr$ (see text for details). The KS probability of the faintest
and brightest PNs being drawn from the same underlying distribution is
only $1.9\%$, while the intermediate PNs are still compatible with the
faint PNs (KS probability $28\%$).}
\label{zezang}
\end{figure}

\begin{figure} 
\includegraphics[width=0.95\linewidth]{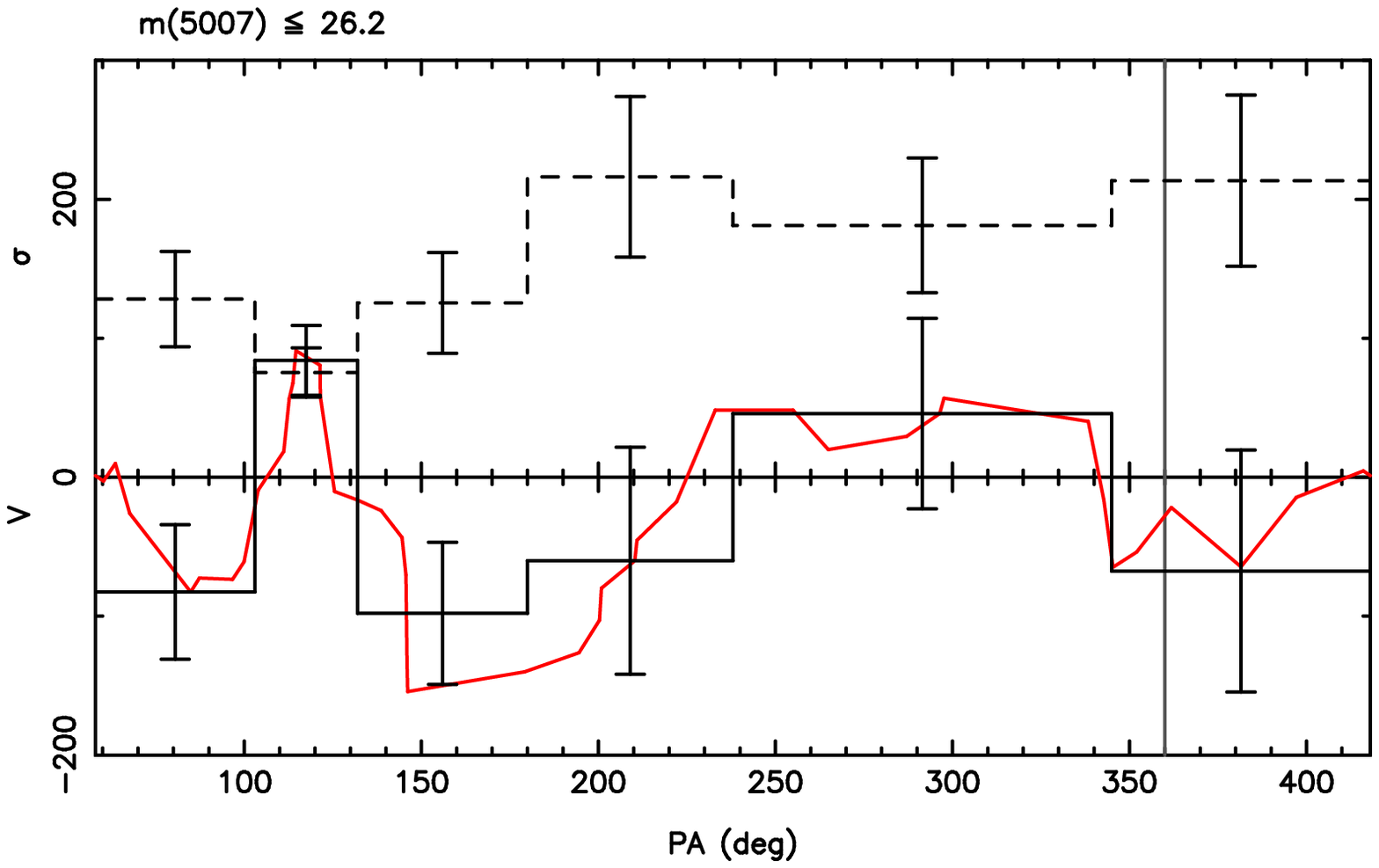}
\includegraphics[width=0.95\linewidth]{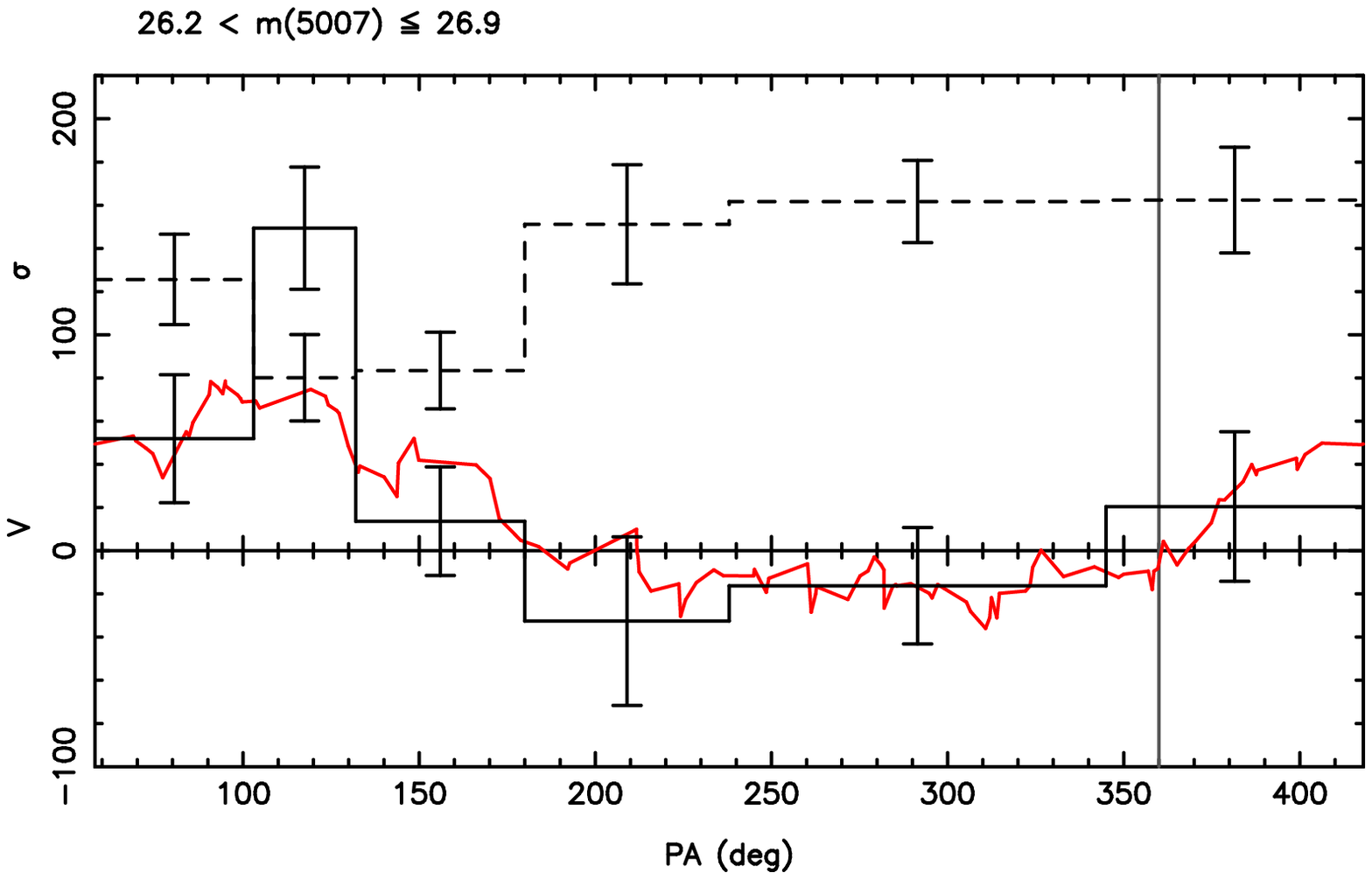}
\includegraphics[width=0.95\linewidth]{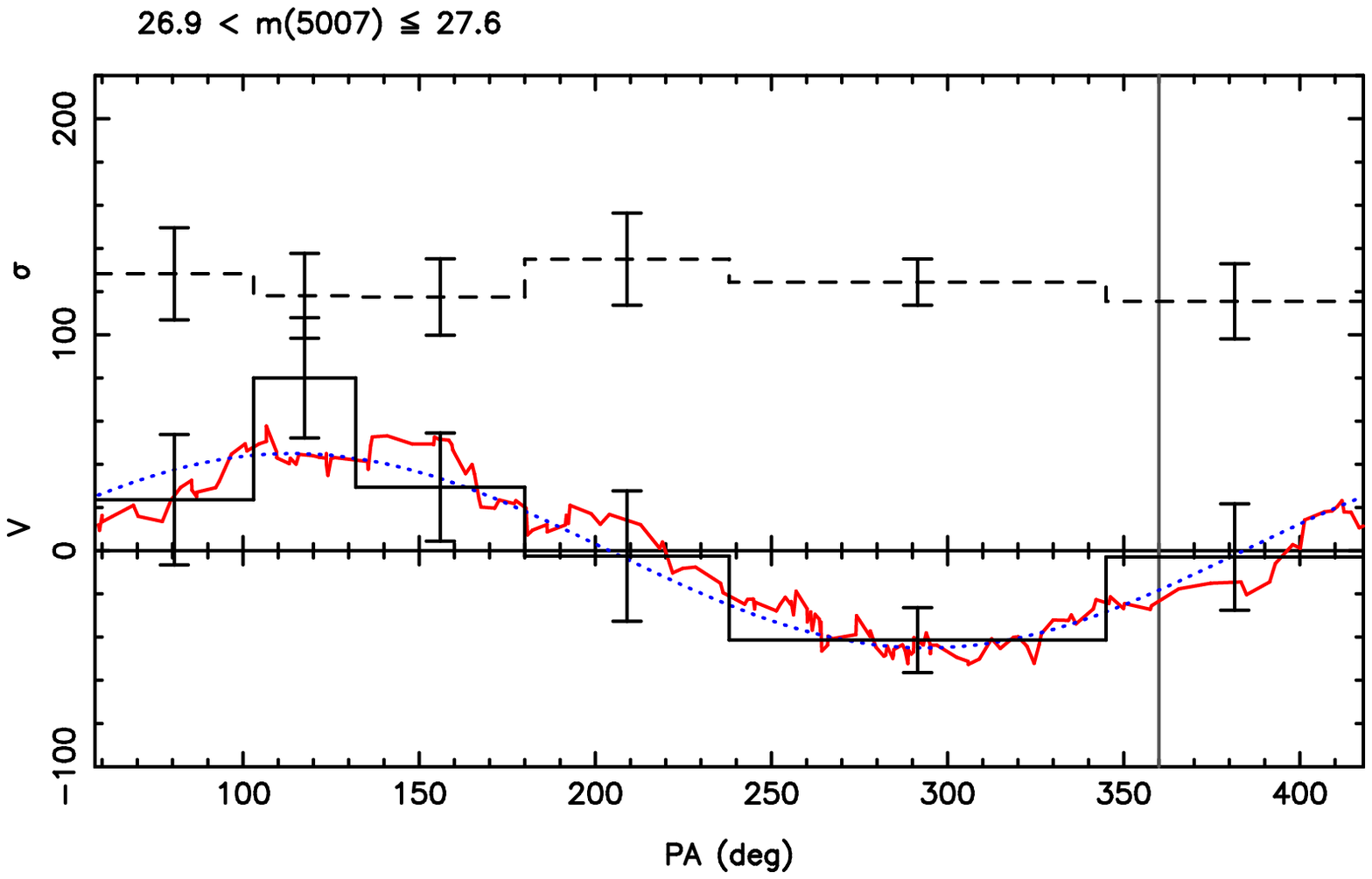}
\caption{Azimuthal distributions of PN velocities. The PA is measured 
clockwise from North, as in the convention of \citet{zez03}, so that
the galaxy major axis is along $\mathrm{PA} = 114\degr,
294\degr$. Solid (dashed) histograms show the mean velocity
(dispersion) and their errors in different angular bins. The red
ragged lines show running averages of mean velocity.
\emph{Top:} Velocities of PNs brighter than $m(5007)=26.2$ in the 
Complete sample. Bin sizes are chosen so as to have approximately
constant number of PNs ($\sim 7$) in each bin. The mean velocity along
the major (minor) axis is positive (negative) on both sides of the
center. The running average is also over $7$ PN velocities. The
velocity dispersion is largest (smallest) on the minor (major)
axis. 
\emph{Middle:} Velocities of PNs with $26.2 < m(5007) \leq 26.9$
in the same angular bins as in the top panel. The running average is
over $27$ PNs.
\emph{Bottom:} Velocities of PNs with 
$26.9 < m(5007) \leq 27.6$ in the same angular bins as in the top
panel. The faint PNs show a rotation pattern as for the absorption
line data but with a smaller peak velocity, and are consistent with a
flat dispersion of $\sim 120 \kms$. The running average is over $33$
PNs. The blue dotted line shows a sinusoidal fit. The kinematics of the
intermediate brightness PNs is intermediate between the faint and
bright PNs.}
\label{vang} 
\end{figure}

The left-right asymmetry is confirmed by inspecting the azimuthal
distribution of the faint and bright PNs. In the literature we found a
related analysis by \citet{zez03} who compared the azimuthal
distribution of \emph{Chandra} X-ray point sources (XPS) with the
optical surface brightness of NGC 4697. We follow their PA convention,
and plot the cumulative angular distributions of the bright,
intermediate, and faint PNs in our Complete sample in
Figure~\ref{zezang}. For comparison, the right panel of Fig.~2 from
\citet{zez03} is also shown. The angular distribution of all PNs in
the Complete sample has a shape somewhere in between that of the XPS
and that of the optical light. The brightest PNs are in complete
disagreement with either of these distributions; they seem to be more
concentrated in a narrow angular sector between $110\degr \la
\mathrm{PA} \la 150\degr$ (see Fig.~\ref{tspace}), with only $1.9\%$
probability that the faint and bright PN subsamples are drawn from the
same azimuthal distribution.  At the same time, the radial
distribution of the faint and bright subsamples are not significantly
different (Figure \ref{radial}).

The velocity distribution of the brightest PNs is also correlated with
their azimuthal distribution. In Figure~\ref{vang} we plot the mean
radial velocity and its dispersion in angular sectors containing
approximately constant number of PNs from the bright subsample, as
well as an angular running average of their mean radial velocity.
Along both sides of the major-axis of the galaxy, the brightest PNs
have positive velocity, while showing a relatively low velocity
dispersion, perhaps due to infall as suggested for the XPS by
\citet{zez03}. Along the minor-axis they have a large dispersion with
a mean velocity that is negative. This kinematics is compatible with
neither the faint PN velocities nor the stellar absorption line data.
On the other hand, the fainter PNs show a regular azimuthal
distribution in the mean line-of-sight radial velocity and velocity
dispersion. Note that their velocity dispersion is in the mean
smaller than that of the bright sample. The kinematics of the
intermediate luminosity PNs is intermediate between those of the
bright and faint subsamples. The intermediate luminosity PNs
thus contain significant contributions from both of the different
populations that dominate the bright and faint bins, respectively.

\begin{figure}
\includegraphics[width=0.92\linewidth]{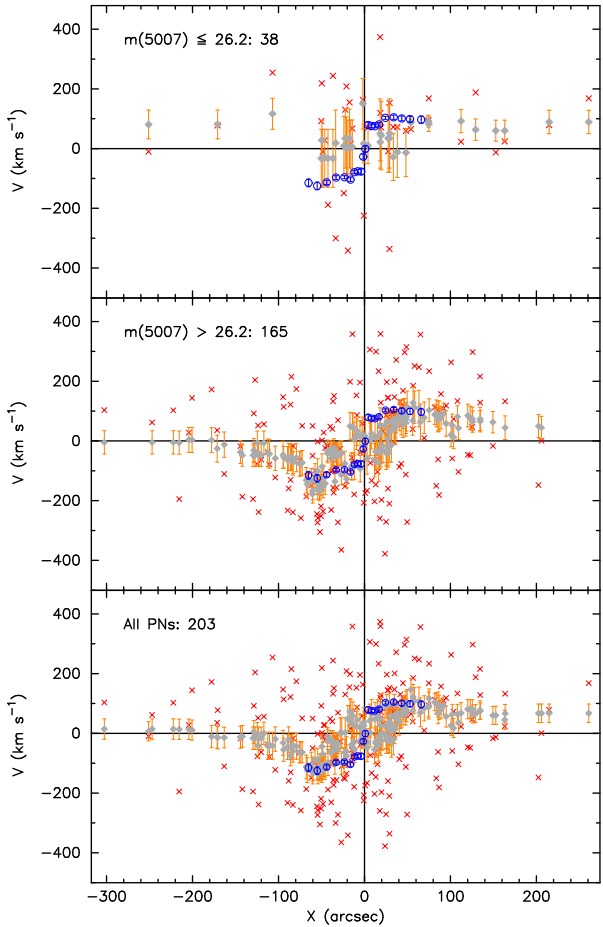}
\caption{Velocity within $\pm 30"$ of the major axis for bright and 
faint/intermediate PNs separated at $m(5007) = 26.2$. Blue circles
show stellar absorption line spectroscopy (ALS) data from
\citet{bin90}, red crosses are individual PN velocities, and grey
diamonds are near neighbour running averages. The faint/intermediate
sample is in approximate agreement with the ALS data, while the bright
sample has a velocity consistent with zero within $\rm{R_{eff}}$ and
positive velocities on both sides of the galaxy center.}
\label{majkindb}
\end{figure}

Furthermore, including the bright PNs in deriving kinematics for the
whole PN sample introduces signficant contamination effects.  In
Figure~\ref{majkindb}, we demonstrate this by plotting the mean PN
velocities along a $\pm 30\arcsec$ wide slit about the major-axis. The PNs
have been separated into faint/intermediate and bright sub-samples at
$m(5007) = 26.2$. Inside the $\rm{R_{eff}}$, the mean velocities of
the faint PNs agree with stellar absorption line spectroscopic (ALS)
data from \citet{bin90}, while the bright PNs have much smaller mean
velocities, consistent with $\sim 0$. In the outer parts, the velocity
asymmetry in the bright PNs leads to a positive mean velocity on both
sides of the galaxy center. Both the faint/intermediate sample and the
entire PN sample also show some asymmetry: the outer mean
velocities on the $X<0$ side reach zero, but not the negative values
expected from reflecting the positive values at $X>0$. This confirms
that also the intermediate/faint subsample contains a fraction of PNs
stemming from the out-of-equilibrium population traced by the bright
PNs.  However, the streaming velocity of the majority of the faint PNs
does appear to decrease on both sides of the center.  Similarly
contaminated results could be expected for the derived velocity
dispersions.

Several important conclusions can be drawn from these figures.  (i)
The bright PNs as defined in Section~\ref{subpop} and
Fig.~\ref{mvplot} do not trace the azimuthal distribution of light in
NGC 4697. (ii) They do not trace the fainter PNs in their azimuthal
kinematics; thus, a large fraction of them must belong to a separate
PN subcomponent originating from a separate stellar population. (iii)
Third, they are not in dynamical equilibrium in the gravitational
potential of NGC 4697. (iv) Because this subpopulation does not trace
the stars, including its PN velocities into dynamical analysis of the
galaxy will lead to significant errors in the results.

\begin{figure}
\includegraphics[width=0.95\linewidth]{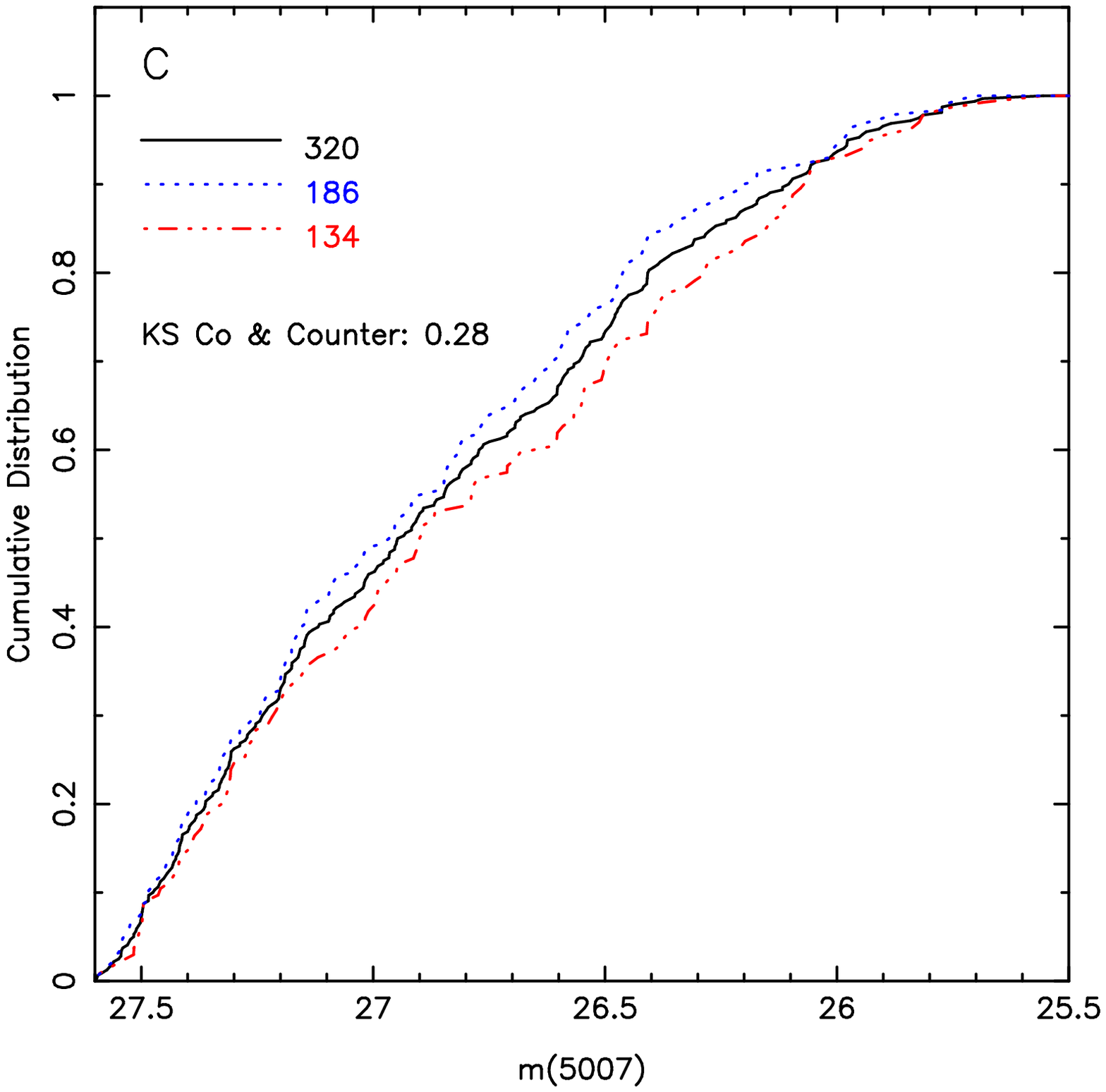}
\includegraphics[width=0.95\linewidth]{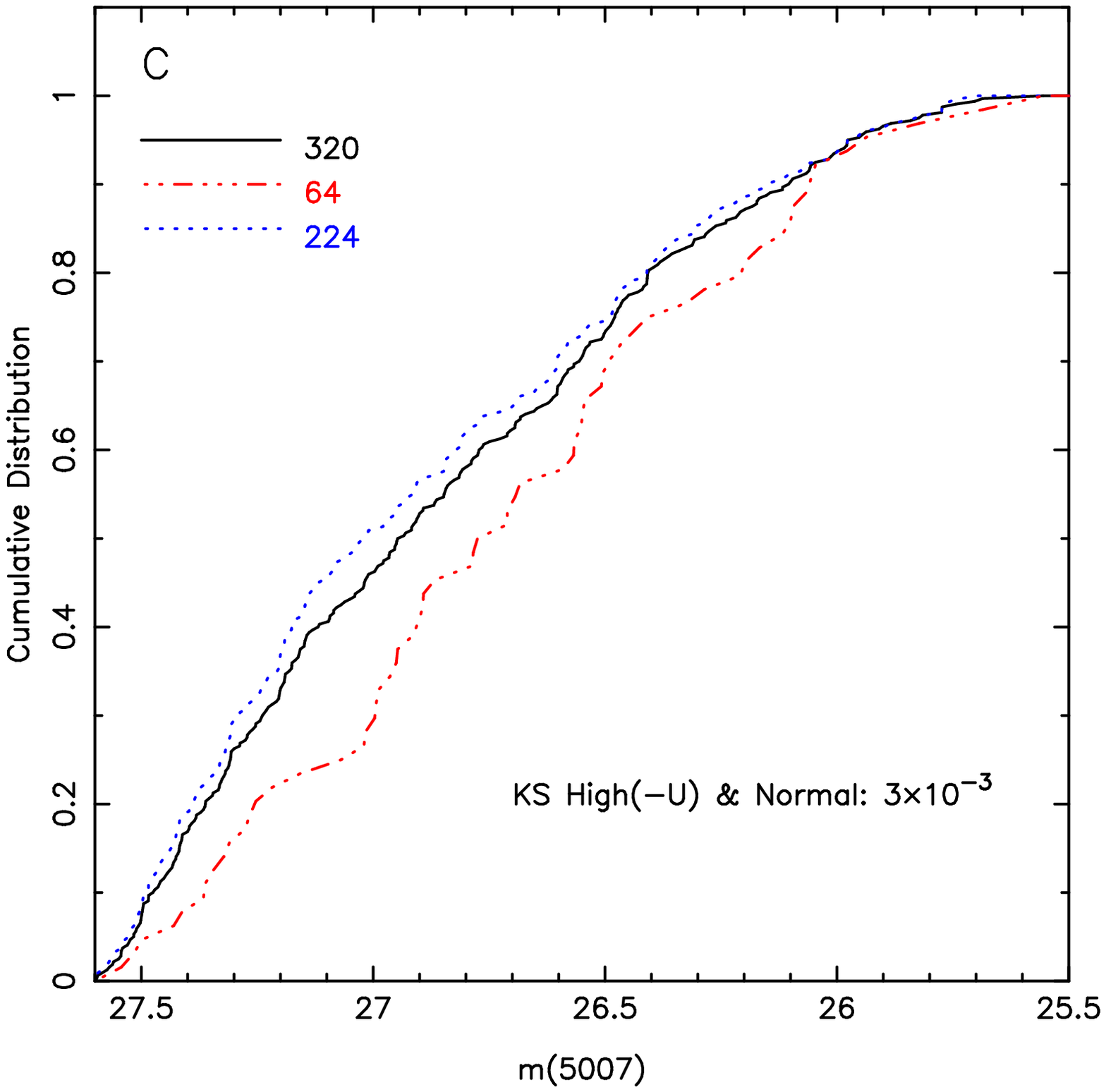}
\caption{\emph{Top:} 
Cumulative distributions of PNs magnitudes for all (black solid line),
co-rotating (blue dotted line) and counter-rotating (red dash-dotted
line) PNs in NGC 4697.  The magnitude distribution of co- and
counter-rotating PNs is similar with only $28\%$
probability. \emph{Bottom:} Cumulative magnitude distributions of the
main PN sample ($-100\kms \le U \le 200\kms$, blue dotted line) and
the extreme counter-rotating sample ($U<-100\kms$, red dash-dotted
line) defined in the text, along with that of the total sample (black
solid line). The first two luminosity functions are different with
99.7\% confidence; thus the PNLF cannot be universal.}
\label{ksmag}
\end{figure}

\subsection{Kinematic dependence of PNLF}

What can we learn about the luminosity functions of the two kinematic
components of the PN system in NGC 4697?  The cumulative magnitude
distributions of the co-rotating and counter-rotating subsamples in
Fig.~\ref{mvplot} have only $28\%$ KS probability of stemming from the
same distribution (top panel of Figure~\ref{ksmag}). However, the fact
that there are both co-rotating and counter-rotating bright PNs in the
bright sample whose azimuthal distribution is unmixed, shows that
counter-rotation is not a {\sl clean} discriminator after all for the
secondary population of PNs in NGC 4697. Thus the luminosity functions
of the main and secondary PN populations in this galaxy may be a lot
more different than this figure of $28\%$ would suggest.

From Fig.~\ref{vang} we estimate that the main population of PNs in
NGC 4697 has a mean radial velocity $\overline{v}_{\rm faint}
\simeq 45\kms \sin({\rm PA}-24\degr)$, with a dispersion of 
$\sigma_{\rm faint} \simeq 120\kms$, so its reduced mean velocity is
$0<\overline{U}_{\rm faint} \lta 45\kms$.
Thus in the bottom panel of Fig.~\ref{ksmag} we show the cumulative
magnitude distribution of the PNs in the velocity range $-100\kms \le
U \le 200\kms$ and compare it with the cumulative distribution of the
PNs with $U<-100\kms$. These velocity ranges are dominated by the main
and secondary PN populations in the sample, respectively.  Now the
Kolmogorov Smirnov significance test shows that the magnitude
distributions from these sections of Fig.~\ref{mvplot} have only
probability $0.3\%$ of being drawn from the same distribution.  This
result is strong enough to imply that the PNLF cannot be universal --
the PNLF in NGC 4697 depends on a kinematic selection.

In the following, we will use the velocity range $-100\kms \le U \le
200\kms$ as an approximate kinematic selection criterion for the main
PN population in NGC 4697.  The luminosity function of the strongly
counter-rotating PNs in Fig.~\ref{mvplot} -- a first approximation to
the luminosity function of the secondary PN population in NGC 4697 --
differs from that of the main PN distribution so defined in the sense
that it contains more bright PNs near the cut-off and fewer faint PNs
than the main population (see Fig.~\ref{ksmag}). Now an important
question is: in what proportion do both populations contribute to the
brightest PNs, and do they have different cutoff magnitudes?

\begin{figure}
\includegraphics[width=0.95\linewidth]{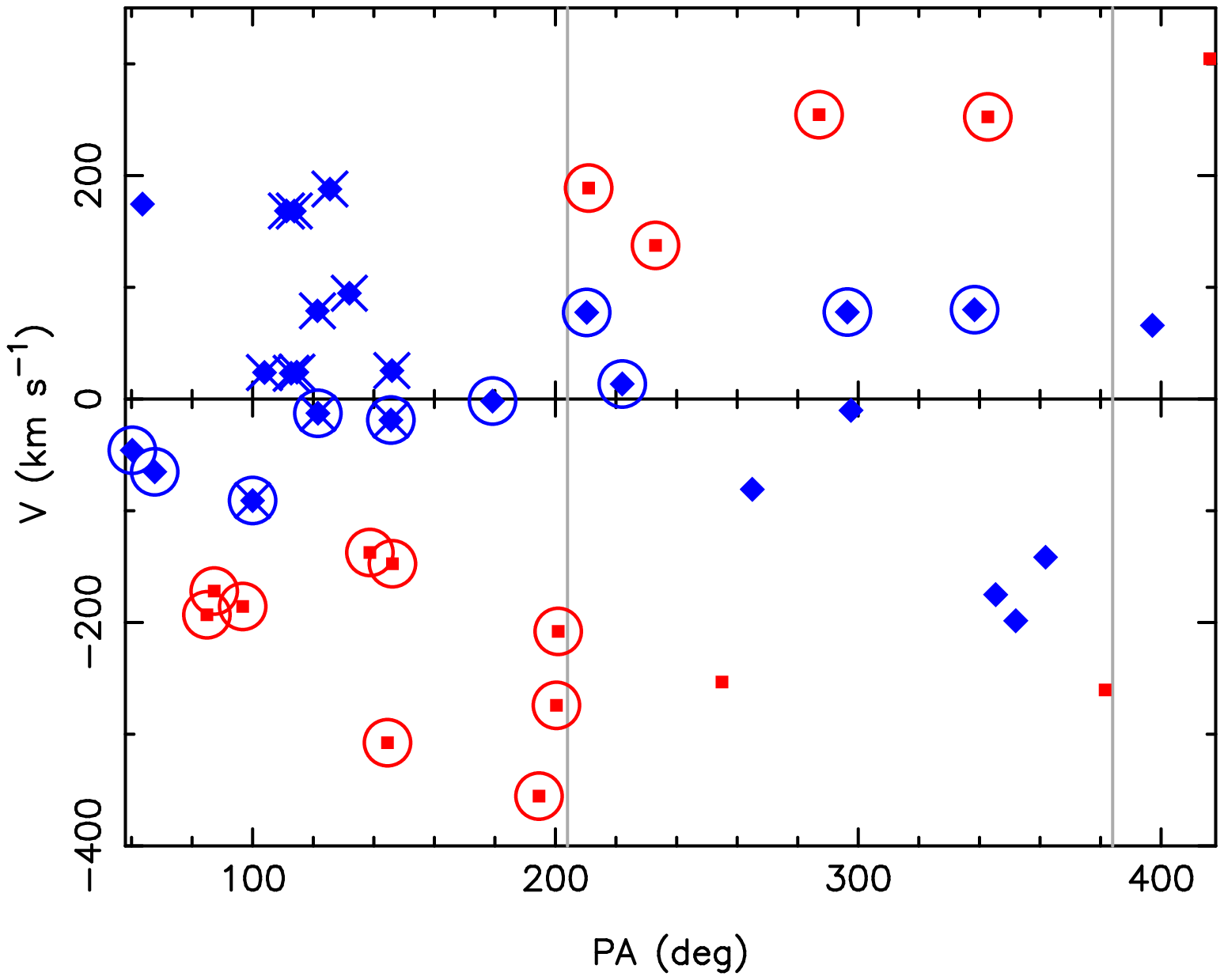}
\includegraphics[width=0.95\linewidth]{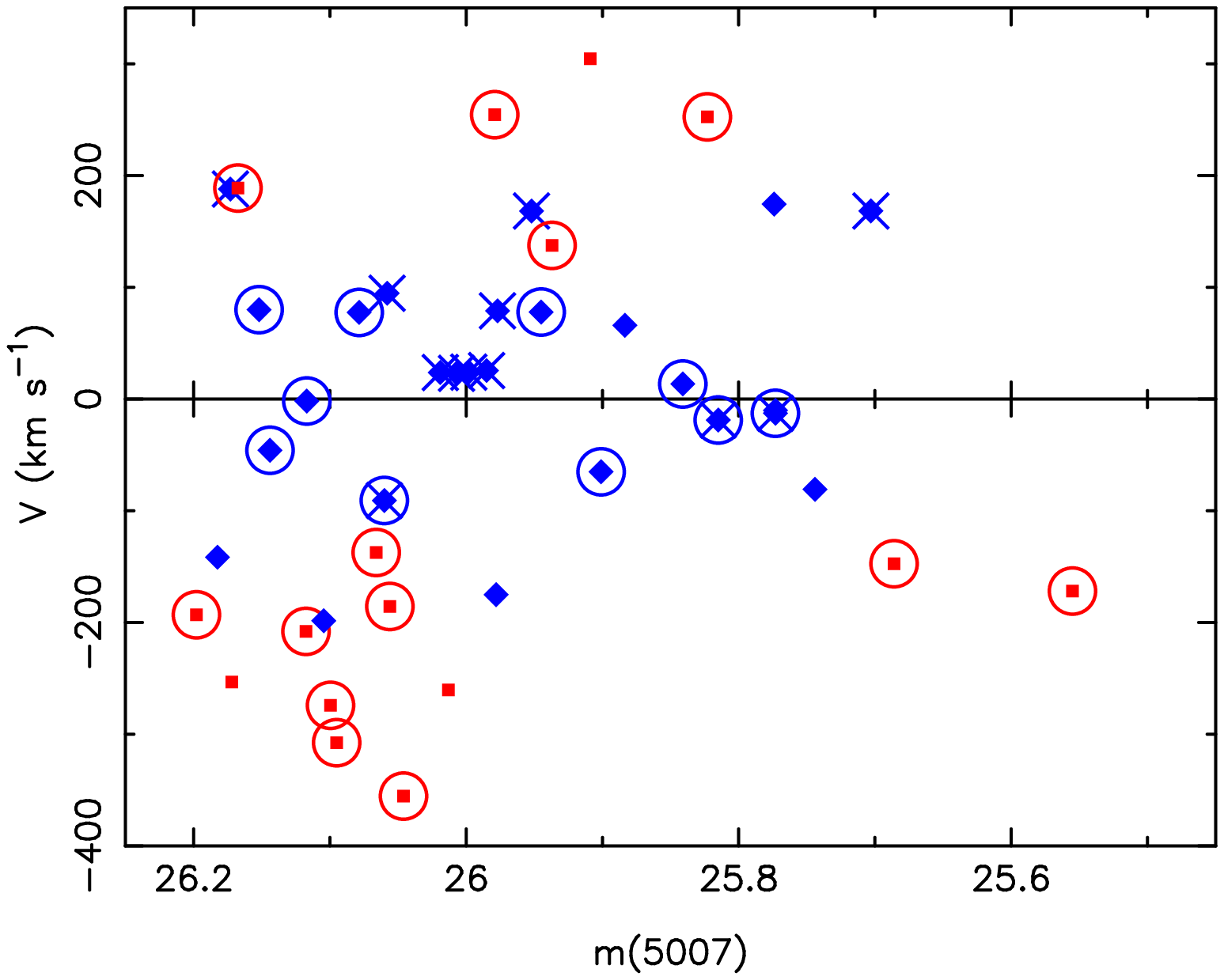}
\caption{Disentangling the bright PN subsample from radial velocities, 
luminosities, and position angles on the sky.
\emph{Top:} Radial velocity -- position angle plane. Red squares (blue
diamonds) show PNs whose reduced velocities are outside (within)
$-100\kms \leq U \leq 200\kms$. Circled symbols denote
counter-rotating PNs; many of these are kinematic outliers. The
two grey lines denote the position angle of the minor axis. The
distribution of blue diamonds is approximatively uniform in PA except
near PA$=100\degr - 150\degr$ where there are 12 PNs (crossed diamonds)
instead of the expected 2 PNs.
\emph{Bottom:} Radial velocity -- magnitude plane. The symbols are as
in the top panel. 6/8 of the brightest PNs are either kinematic outliers
or are found in the overdense angular region. 
}
\label{vla}
\end{figure}

To investigate this, we show in Figure~\ref{vla} the velocities,
magnitudes, and position angles of the entire bright PN subsample (see
Figs.~\ref{mvplot} and \ref{vang}). The top panel shows (i) that 13/16
of the bright PNs, whose radial velocities differ most from those of
the faint population, are counterrotating. This explains the
differences between the velocity distributions of co- and
counter-rotating PNs that first suggested more than one PN population
in Section~\ref{subpop}.  (ii) Also, even in the kinematically {\rm
normal} bright PNs, there is a large overdensity (10/12) in the
angular range PA$=100\degr - 150\degr$.  The bottom panel shows in
addition that many of the brightest PNs are either kinematic outlyers
or found in the angular overdensity (6/8). 

\begin{figure}
\plotone{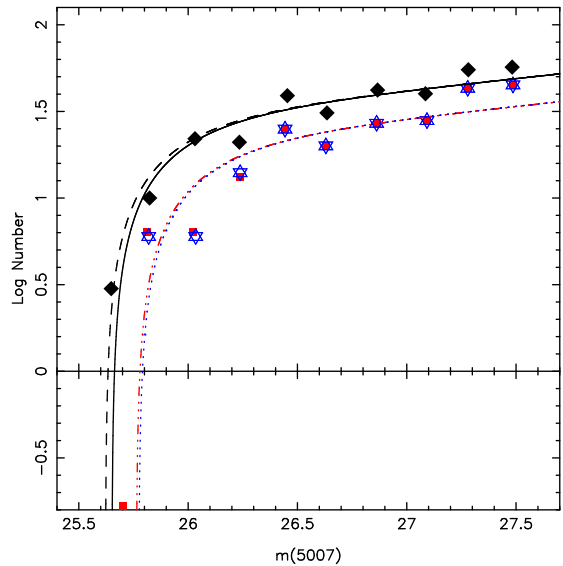}
\caption{Distributions of PN magnitudes in NGC 4697. The PNLF for 
all PNs (black diamonds, dashed black line) is similar to the PNLF of
\citet{men01} (black solid line).  The red squares and $\chi^2$-fitted
dash-dotted line show the PNLF of the main population of PNs in the
velocity range $-100\kms \le U \le 200\kms$, but with the bright PNs
in the angular overdensity weighted only by $2/12$ so that the angular
distribution of the bright PNs becomes approximately uniform. The blue
stars and $\chi^2$-fitted blue dotted line show the PNLF in the same
velocity range, but now with the 10 brightest PNs in the angular
overdensity removed. }
\label{clumfun}
\end{figure}

Clearly, to arrive at a main population of NGC 4697 PNs that is in
dynamical equlibrium in the gravitational potential, we must remove
the angular overdensity. Then we are left with 16 PNs in the range
$-100\kms \leq U \leq 200\kms$ of a total bright subsample of 42 PNs;
however, there is some freedom in the way the angular overdensity is
removed. In the following, we explore two assumptions: (i) using all
12 PNs in the angular overdensity, but weighting each one by
$2/12=0.17$, and (ii) removing the brightest 10 of the 12 PNs in the
overdensity.  The resulting luminosity functions for both cases are
plotted in Figure~\ref{clumfun}; they differ only slightly. 

Thus in the following we use the kinematic condition $-100\kms \leq U
\leq 200\kms$ together with  assumption (ii) above to ensure azimuthal 
uniformity of the bright PNs, as an improved selection criterion for
PNs in the main population in NGC 4697. We keep all PNs in the
intermediate luminosity bin with $-100\kms \leq U \leq 200\kms$,
because Fig.~\ref{zezang} showed that their azimuthal asymmetry is not
large. Also, we have checked that the mean angular velocity distribution
in this magnitude bin after applying the kinematic selection follows
approximately the sinusoidal variation of the faint PNs, and the
velocity disperion is approximately constant.  The resulting main
population sample is identified by their brightness, distribution and
kinematic properties.

Comparing the luminosity function in Figure~\ref{clumfun} of this main
PN population, with the luminosity function of all PNs in
NGC 4697, we see that the bright cutoff of the main population is
shifted to fainter magnitudes. We note that the bright cutoff could be
shifted further to fainter magnitudes if some of the kinematically
normal and azimuthally uniform bright PNs were also part of the
secondary population, which we cannot tell from the present data.

\begin{figure}
\includegraphics[width=0.95\linewidth]{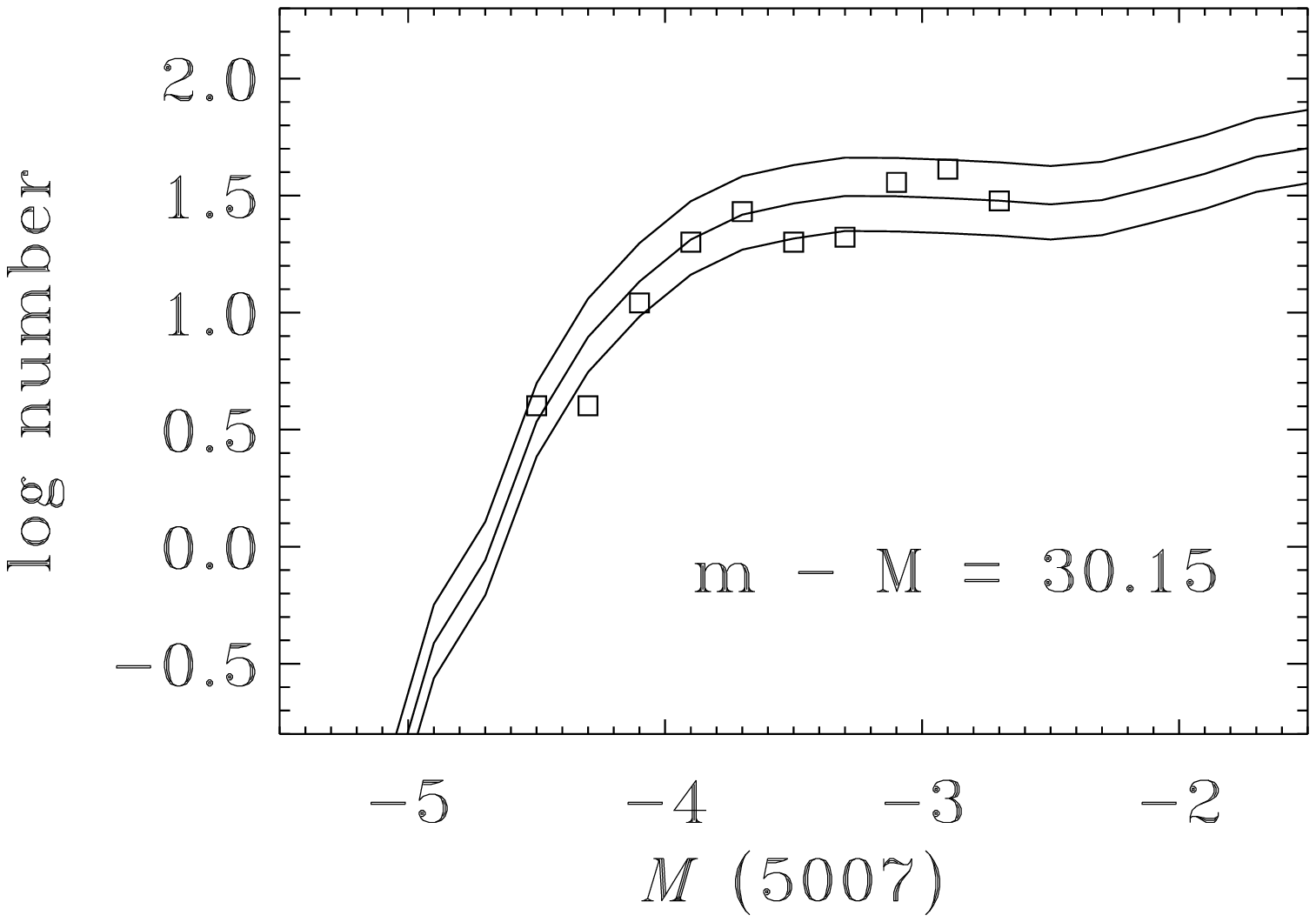}
\includegraphics[width=0.95\linewidth]{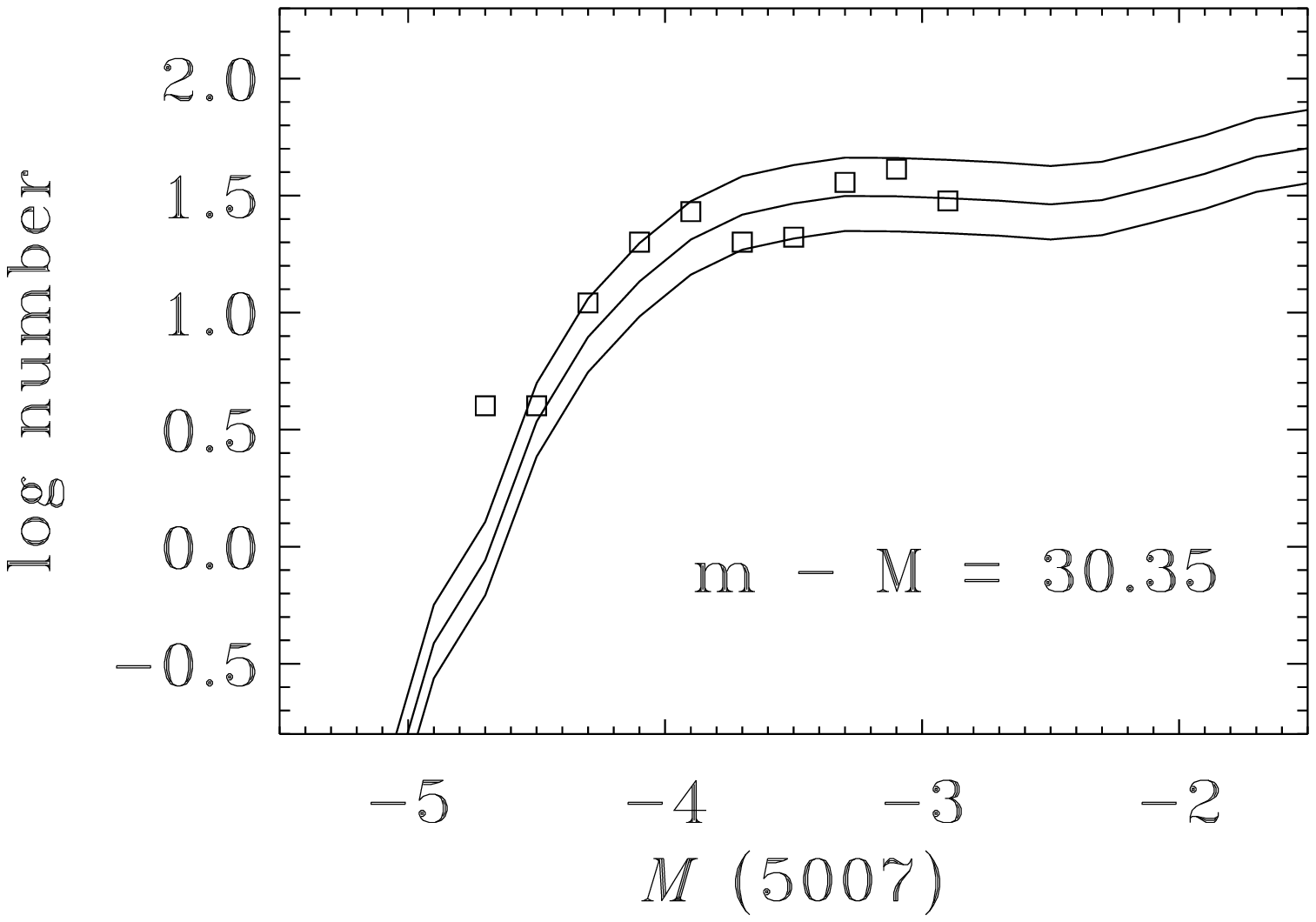}
\caption{
PNLF of NGC 4697 (squares), built from the main population PN sample
(blue stars in Fig.~\ref{clumfun}) of 214 objects, binned into 0.2 mag
intervals. The apparent magnitudes $m(5007)$ have been transformed
into absolute magnitudes $M(5007)$ by adopting an extinction
correction of 0.105 mag and distance moduli indicated in each
plot. The three lines are PNLF simulations \citep{men97}. For a
distance modulus 30.15 the brightest PNs are a bit too weak, therefore
the distance must be increased. But 30.35 is clearly excessive. The
best fit is for 30.2 or 30.25.  }
\label{rob}
\end{figure}

We can ask now what is the effect, in practice, on the PNLF distance
determination. The reduced sample for the main PN population in NGC
4697 has 214 objects. After binning these data into 0.2 mag intervals,
we transform the apparent magnitudes $m(5007)$ into absolute
magnitudes, adopting the extinction correction of 0.105 mag
\citep{men01} and assuming different distance moduli, and we compare
the results with the PNLF simulations of \citet{men97}. This is the
same procedure used in \citet{men01} for the PNLF distance
determination. The comparison is shown in Fig.~\ref{rob}. We conclude
that the PNLF distance modulus should be increased slightly from 30.1
(the earlier determination based on the full sample) to 30.2 or
30.25. For comparison, the $\chi^2$-fit to the same data (blue line
in Fig.~\ref{clumfun}) gives 30.22. This correction would bring the 
PNLF distance modulus in better (but not perfect) agreement with 
the surface brightness fluctuation (SBF) distance modulus ($30.35\pm0.2$) 
reported by \citet{tonry}.

\section{Discussion}
\label{discn}

\subsection{Can the NGC 4697 PN sample be contaminated?}

Bright PNs play a significant role in the analysis of the last
section. Is it possible that the brightest PNs in NGC 4697 are
contaminated by compact HII regions such as those observed in
\citet{ger02, ryweb04}? The recent observations of \citet{men05} 
have shown that this cannot be so. These authors took spectra of 13/42
PNs in our bright subsample in NGC 4697 with FORS2{@}VLT; these have
no detectable continuum and the line ratios of metal-rich PNs.

The same argument also shows that these bright PNs cannot be
background emission line galaxies. Moreover, Ly alpha emission
galaxies would come in at [OIII] magnitudes of m$_{5007}\gta 26.2$
\citep[see Fig.~4 of][]{agu05}, while the bright PNs in NGC 4697 have
m$_{5007}\le 26.2$.

Furthermore, one may wonder whether the bright PNs in NGC 4697 might
simply be foreground objects which could be closer to us and hence
brighter than the 'true' NGC 4697 PNs on which they would be superposed.
Given that NGC 4697 is located in the southern extension of the Virgo
cluster, known to contain an intracluster population of PNs
\citep{arn02,arn04,fel04}, this possibility deserves to be considered. 
However, the following observational facts show that the bright PNs in
NGC 4679 are not intracluster PNs (ICPNs). (1) NGC 4697 is in fact
closer than the Virgo cluster. The PNLF from \citet{men01} places it
at $10.5\pm 1\mpc$ (m-M$= 30.1$), while the distance modulus from the
PNLF of M87 is m-M$= 30.7\pm 0.1$ \citep{cia02}. Even if we discarded
the entire brightest 0.3 mag of the NGC 4697 PNs, NGC 4697 would still
be at 75\% of the distance of M87. (2) The velocity dispersion of the
bright and unrelaxed PNs in NGC 4697 is $\sim 170\kms$, but varying
azimuthally, while that of the fainter main population is $123\kms$
(Table \ref{tabvtftest}, Figs.~\ref{mvplot}, \ref{zezang}). Both are
much smaller than the velocity dispersion of ICPNs in Virgo
\citep{arn04}.  The radial distribution of the bright population
outside the incompleteness ellipse is concentrated towards the galaxy
center and KS compatible (92\%) with the radial distribution of the
fainter PNs (see Fig.~\ref{radial}). Thus also the bright PNs in NGC
4697 are bound to the galaxy. (3) The surface density of PNs with
m$_{5007}\le 26.2$ is 0.58 PNs/arcmin$^2$, while the mean surface
density of Virgo ICPNs is 0.02 PNs/arcmin$^2$, much smaller
\citep{agu05}. This last argument also rules out significant
contamination by chance superpositions of PNs even closer than NGC
4697.

We conclude that the bright PN population in NGC 4697 consists of
genuine PNs, and that it is dynamically bound to the NGC 4697
system. The irregular angular distribution and kinematics of these
PNs, by comparison with the fainter main population, show that they
must belong to a separate stellar population not yet in dynamical
equilibrium with NGC 4697.

\subsection{Comparison with other galaxies}

PN samples as large as the one for NGC 4697 are still the exception.
Yet to undertake the analysis described in Section~\ref{analysis} it
was crucial to work with a complete sample of some 300 PNs. The PN.S
project \citep{dou02} may lead to complete samples of similar size but
as of now the typical sample sizes are $\lta 200$ \citep{rom03} and
their magnitude distributions and completeness have not been
studied. The only other elliptical galaxy with a comparable (even
larger) sample is Cen A \citep{hui95,pen04}. A detailed analysis of
the distribution of Cen A PNs in the magnitude--velocity plane is
still pending, but \citet{hui93} analysed the PNLF in Cen A as a
function of radius, based on narrow-band surveys. They concluded that
no population effect on the PNLF bright cut-off could be seen,
suggesting that the 0.3 mag difference between their main and outer
halo samples was due to filter transmission uncertainties. Analysis
of the large sample from \citet{pen04} will clarify whether the PNs in
Cen A are consistent with one or more subpopulations. This will be
particularly interesting because Cen A is believed to be the remnant
of a galaxy merger, so one might expect PNs from both the older
stellar populations of the progenitors as well as from the stars
formed in the subsequent interaction between them.

Without large kinematic samples, searches for PN subpopulations must
be based on the PN luminosity distributions. The work of Jacoby,
Ciardullo and collaborators cited in the Introduction has shown that
the PNLF is remarkably uniform. However, there are exceptions: we
recall that in the halo of M84 there exists a small population of
overluminous PNs whose cutoff is $0.3$ mag brighter than that of the
main M84 population, but which appear nonetheless bound to the halo of
M84 \citep{arn04,agu05}. They must therefore be intrinsically bright,
due to some stellar population difference.  In M87, the only
overluminous PNs projected onto the galaxy for which velocities have
been measured \citep{arn04}, have very large relative velocities with
respect to M87.  This is most naturally explained if these PNs have
fallen into the deep potential well of M87 from far out in the
cluster; this would again imply an intrinsic population
difference. However, a larger kinematic sample in M87 is required to
be certain.

The existence of the bright PN subpopulation in NGC 4697 implies some
uncertainty in the PNLF cut-off luminosity for this galaxy.  Depending
on whether or not the azimuthally symmetric brightest PNs are part of
the main population, the cut-off luminosity of the main PN population
in NGC 4697 is fainter than that of the whole population by $\gta0.15$
mag (Figs.~\ref{clumfun}, \ref{rob}).  While this is consistent with
the M84 result, is it also consistent with the systematic studies of
PNLF distances? \citet{cia02} have compared the PNLF and SBF
distances, finding a distribution of residuals with a systematic
offset by $\simeq 0.3$ mag, which they suggested is due to internal
extinction effects, and with a FWHM of $\simeq 0.5$ mag. The offset
has in the mean-time been reduced to $\simeq0.15$ mag, following a
revision of the SBF distance scale by \citet{jens03}.  The width of
this distribution is consistent with their determination of the
observational errors in both methods. However, the offset we have
determined for NGC 4697 is also consistent with the distribution of
residuals in \citet{cia02}.

\subsection{Origin of the PN population difference: a 
secondary, younger stellar population?}

We have shown that a large fraction of the bright PNs in NGC 4697
belong to a secondary, dynamically young stellar population that is
not well-mixed in the gravitational potential of the galaxy. Late
infall of tidal structures \citep{zez03} or a merger with a smaller
galaxy some time ago would be natural ways to add such an unmixed
stellar component to NGC 4697. What physical population difference is
correlated with this dynamical youth?

\citet{men05} show from their spectroscopic data for 13
bright PNs that these have near-solar metallicities. Of these 13
bright PNs, 6 are inside the incompleteness ellipse, one has no
measured velocity, and 6 belong to our secondary population.
\citet{men05} also use long-slit spectroscopy to show that the 
metallicity of the integrated stellar population within one effective
radius has solar or higher metallicities. These observations make it
unlikely that metallicity is the main factor responsible for the
different magnitude distributions of the main and secondary PN
populations in NGC 4697.

Thus the more likely driver would appear to be an age difference, as
suggested by \citet{mar04} and as might generally be expected in an
accretion event. Based on their result that the distribution of X-ray
point sources in NGC 4697 does not follow the stellar light,
\citet{zez03} have argued that this is because these sources were
formed several $10^8$ years ago in tidal tails that are now falling
back onto the galaxy. Note, however, that the integrated light in NGC
4697 shows no evidence of young stars with mean age $<7$ Gyr
\citep{men05}, so this younger component could not be luminous enough
to contaminate the integrated light to the level measured.  Also note
that the observed increase of extinction in the PN envelope with PN
core mass more than compensates for the increase of core luminosity
with core mass, for bright PNe in local group galaxies
\citep{cia99}, so that stars with ages below $1$ Gyr may not reach the
[OIII] luminosity at the PNLF cutoff. A secondary stellar population
younger than $1$ Gyr is therefore unlikely as well.

Recently, \citet{cia05} have argued that the brightest PNs in the PNLF
must have core masses of $\gta 0.6\msun$, corresponding to main
sequence masses of $\sim 2.2\msun$. They argue further that for such
high-mass objects to occur in elliptical galaxies, these early-type
galaxies would either have to contain a small, smoothly distributed
component of young ($\lta 1$ Gyr age) stars, or more likely, that the
bright PNs in these systems have evolved from blue straggler stars
created through binary evolution. Their blue straggler model, due to
the assumption of a fixed distribution of primary-to-secondary mass
ratios for the initial binaries, predicts that older stellar
populations produce fewer bright PNs per unit luminosity, as is
observed, because the number of binary stars in a stellar population
that can coalesce to $\sim 2.2\msun$ blue stragglers decreases with
time.

If correct, this blue straggler model could also explain how the
secondary population we found in NGC 4697 can contain a large fraction
of the brightest PNs in this galaxy, provided that the stellar
population corresponding to this secondary PN population is
appreciably younger than the main stellar population, whose age is
$\sim 9$ Gyr from optical spectroscopy \citep{tra00}. At the same
time, this secondary stellar population must not be so young to
violate either the constraints from the optical colours or from the
envelope absorption - PN core mass correlation, i.e., must be older
than $\sim 1$ Gyr.  We can give an estimate for the effect of such an
intermediate age population on the optical colours as follows. The
secondary subpopulation traced by the bright and predominantly
counterrotating PNs may contain $\sim 20\%$ of all PNs in the Complete
sample for NGC 4697. A stellar population as blue as the bulge of M31
has a luminosity-specific PN density per unit $L_B$ up to 5 times
higher than the populations characteristic for old elliptical galaxies
\citep{hui93,cia05}. Thus the subpopulation corresponding to the
secondary PN population in NGC 4697 would be expected to contain $\gta
5\%$ of the blue luminosity of NGC 4697, spread over a large fraction
of at least the E image. To detect this we need deep and accurate
photometry.

The unmixed spatial and velocity distributions of the secondary PN
population in NGC 4697 show that that this population is {\sl
dynamically young}, i.e, has not had time to phase-mix and come to
dynamical equilibrium in the gravitational potential of NGC 4697. It
may well be associated with tidal structures that were formed in a
merger or accretion event $\sim 1-2$ Gyr ago, and that have now fallen
back onto the galaxy, or be associated with a more recent
merger/accretion with a red galaxy such as described in
\citet{vdokk05}. In a Universe in which structures form
hierarchically, such secondary stellar populations might be quite
common in ellipitical galaxies, but they would be difficult to
see. The present work shows that studying their PN populations is one
promising approach of looking for such secondary populations. However,
large PN samples are required; most existing PN studies of early-type
galaxies do not have the statistics for such an investigation.
Moreover, in only a fraction of cases may there be enough asymmetry
signal to detect with a few hundred PNs.

\section{Conclusions and Implications}
\label{result}

We have analysed the magnitudes, kinematics and positions of a
complete sample of 320 PNs in the elliptical galaxy NGC 4697 from
\citet{men01}. This data set is large enough for drawing statistically
significant conclusions, and it does not suffer from detection
incompletenesses in either magnitudes or radial velocities. 
We know of no systematic effects in the data that could explain
our results. Our main conclusions are:

\begin{enumerate}

\item Bright and faint PNs in NGC 4697 have significantly different radial 
velocity distributions. The mean velocities of the faint and bright
subsamples (co-rotating and $\sim 0$, respectively) and their velocity
dispersions are different, with 94\% and 99.3\% confidence. Thus the
PNs in NGC 4697 do not constitute a single population that is a fair
tracer of the distribution of all stars.

\item The luminosity functions of the extreme counter-rotating subsample
($U<-100\kms$) and of the main population (defined by $-100\kms \le U
\le 200\kms$) are different with 99.7\% confidence.  The PNLF is
therefore not universal.

\item Based on this, we suggest that there exist (at least) two
PN populations in this galaxy. The secondary PN population in NGC
4697 is prominent in a sub-sample of counter-rotating PNs brighter
than $26.2$. The luminosity function of the entire extreme
counterrotating sample may be a first approximation to the luminosity
function of the secondary population.

\item The spatial distribution of bright PNs with $m(5007) <
26.2$ is different from that of the faint PNs. The bright PNs do not
follow the azimuthal distribution of the optical light, show a
left-right asymmetry, and have a positive mean radial velocity on both
sides of the galaxy major axis, but zero velocity and larger
dispersion on the minor axis. They are not in dynamical equilibrium in
the potential of the galaxy. The fainter population has rotation
properties more similar to the absorption line velocities, with
azimuthally constant dispersion.

\item Using both their kinematics and angular distribution,
we can estimate a lower limit to the statistical fraction of bright
PNs in the secondary population. Based on this we estimate that the
bright cutoff of the main population is uncertain by $\sim 0.15$
mag.

\end{enumerate}

Our results have two main implications for the use of PNs in
extragalactic astronomy. First, for distance determinations with the
PNLF, it may be important to understand how uniform the PN populations
in the target galaxies are. From our analysis in NGC 4697 we estimate
that unrecognized subpopulations of PNs in smaller samples than that
in NGC 4697 may lead to variations of $\sim 0.15$ mag in the bright
cutoff. This would correspond to distance errors of some $10\%$,
which, although a minor effect, could be significant in some cases. It
will be necessary to verify how frequently such subpopulations occur
in elliptical galaxies. We recall that also in the halo of M84 there
exists a small population of overluminous PNs whose cutoff is $0.3$
mag brighter than that of the main M84 population, but which appear
nonetheless bound to the halo of M84 \citep{arn04,agu05}.

The second implication concerns the use of PNs as tracers for the
angular momentum and gravitational potentials of elliptical galaxies.
Our analysis has shown that in NGC 4697 the bright PNs do not trace
the distribution and kinematics of stars and are not in dynamical
equilibrium in the gravitational potential of the galaxy. The fainter
PNs look more regularly distributed but may also contain a fraction of
PNs that belongs to this out-of-equilibrium population. Clearly
therefore, mass determinations based on PN kinematics will in future
require careful study of the PN samples being used, not only to verify
that these PNs are in dynamical equilibrium, but also to test for
different dynamical components. Even if in equilibrium, a younger
population of stars may be more flattened or have a steeper fall-off
than the main body of the elliptical galaxy. If the PNs from this
population are indeed somewhat brighter than the main population, one
can recognize such differences from their lower velocity dispersion or
different radial density profile. However, deep observations and large
PN samples will be required.

\acknowledgments

We are grateful to M.~Arnaboldi, R.~Ciardullo and R.~Saglia for
helpful discussions, and to K.~Freeman, G.~Jacoby, E.~Peng and
A.~Romanowsky for helpful comments on the manuscript. NS and OG thank
the Swiss Nationalfonds for financial support under grant
200020-101766. RHM would like to acknowledge support by the U.S.\
National Science Foundation, under grant 0307489.


\begin{thebibliography}{}

\bibitem[Aguerri \etal (2005)]{agu05} Aguerri, J.A.L. \etal\  
2005, \aj, 129, 2585

\bibitem[Arnaboldi \etal (1998)]{arn98} Arnaboldi, M. \etal\ 1998, \apj, 
507, 759

\bibitem[Arnaboldi  \etal (2002)]{arn02} Arnaboldi,  M.~et al.\  2002, \aj, 
123, 760

\bibitem[Arnaboldi \etal (2004)]{arn04} Arnaboldi, M., Gerhard, O.E., Aguerri, 
J.A.L., Freeman, K.C., Napolitano, N., Okamura, S., Yasuda, N., et al.\ 2004, 
ApJ, 614, L33

\bibitem[Binney, Davies \& Illingworth(1990)]{bin90} Binney, J., Davies, R.L.,
\& Illingworth, G.D. 1990, \apj, 361, 78

\bibitem[Carter(1987)]{car87} Carter, D. 1987, \apj, 312, 514

\bibitem[Ciardullo \& Jacoby(1999)]{cia99} Ciardullo, R., \& Jacoby, G.H.
1999, \apj, 515, 191

\bibitem[Ciardullo(2003)]{cia03} Ciardullo, R. 2003, in Stellar Candles
for the Extragalactic Distance Scale, ed. D. Alloin \& W. Gieren, Lect
Notes Phys., 635, 243

\bibitem[Ciardullo \etal (2002)]{cia02} Ciardullo, R. \etal\ 2002, \apj, 
577, 31

\bibitem[Ciardullo \etal (1989)]{cia89} Ciardullo, R., Jacoby, G.H., Ford, 
H.C., \& Neill, J.D. 1989, \apj, 339, 53

\bibitem[Ciardullo \etal (2005)]{cia05} Ciardullo, R., Sigurdsson S.,
Feldmeier, J.J., \& Jacoby, G.H. 2005, \apj, 629, 499

\bibitem[Dejonghe \etal (1996)]{dej96} Dejonghe, H. \etal\ 1996, \aap, 306, 
363

\bibitem[Dekel \etal (2005)]{dek05} Dekel, A., Stoehr, F., Mamon, G.A.,
Cox, T.J., Novak, G.S., \& Primack, J.R. 2005, Nature, astro-ph/0501622

\bibitem[Dopita \etal (1992)]{dop92} Dopita, M.A., Jacoby, G.H., \&
Vassiliadis, E. 1992, \apj, 389, 27

\bibitem[Douglas \etal (2002)]{dou02} Douglas, N.~ \etal\ 2002, PASP, 114, 1234

\bibitem[Feldmeier,  Ciardullo, \&  Jacoby (1998)]{fel98}  Feldmeier, J.~J., 
Ciardullo, R., \& Jacoby, G.~H.\ 1998, \apj, 503, 109 
 
\bibitem[Feldmeier \etal (2004)]{fel04} Feldmeier, 
J.~J., Ciardullo, R., Jacoby,  G.~H., \& Durrell, P.~R.\ 2004, \apj,
615, 196
 
\bibitem[Ferrarese \etal (2000)]{fer00} Ferrarese, L. \etal\ 2000, \apj, 
529, 745

\bibitem[Gerhard \etal (2002)]{ger02} Gerhard, O. \etal\ 2002, \apj, 
580, L121

\bibitem[Gerhard \etal (2005)]{ger05} Gerhard, O., Arnaboldi, M., 
Freeman, K.C., Kashikawa, N., Okamura, S., \& Yasuda, N., 2005, 
\apjl, {621}, {L93}

\bibitem[Goudfrooij \etal (1994)]{gou94} Goudfrooij, P., \etal\ 1994, AAS,
104, 179

\bibitem[Hui \etal (1993)]{hui93} Hui, X., Ciardullo, R., \& Jacoby, G.H. 
1995, \apj, 414, 463

\bibitem[Hui \etal (1995)]{hui95} Hui, X., Ford, H.C., Freeman, K.C., \&
Dopita, M.A. 1995, \apj, 449, 592

\bibitem[Jacoby(1997)]{jac97} Jacoby, G.H. 1997, in The Extragalctic 
Distance Scale, Space Telescope Science Institute Series, ed. M. Livio 
(Cambridge University Press), 197

\bibitem[Jacoby(1989)]{jac89} Jacoby, G.H. 1989, \apj, 339, 39

\bibitem[Jacoby, Ciardullo, Ford(1990)]{jcf90} Jacoby, G.H., Ciardullo, 
R., \& Ford, H.C. 1990, \apj, 356, 332

\bibitem[Jensen \etal (2003)]{jens03} Jensen, J.B., \etal 2003, 
\apj, 583, 712

\bibitem[Marigo \etal (2004)]{mar04} Marigo, P., Girardi, L., Weiss, A., 
Groenewegen, M.A.T., \& Chiosi, C. 2004, \aap, 423, 995

\bibitem[M\'endez(1999)]{men99} M\'endez, R.H. 1999, in Post-Hipparcos 
Cosmic Candles, ed. A. Heck \& F. Caputo (Dordrecht: Kluwer), 161

\bibitem[M\'endez \& Soffner(1997)]{men97} M\'endez, R.H., \& Soffner, T. 
1997, \aap, 321, 898

\bibitem[M\'endez \etal (1993)]{men93} M\'endez, R.H., Kudritzki, R.P., 
Ciardullo, R., \& Jacoby, G.H. 1993, \aap, 275, 534

\bibitem[M\'endez \etal (2001)]{men01} M\'endez, R.H. \etal\ 2001, 
\apj, 563, 135

\bibitem[M\'endez \etal (2005)]{men05} M\'endez, R.H. \etal\ 2005,
\apj, 627, 767

\bibitem[Merritt \& Saha(1993)]{ms93} Merritt, D., \& Saha, P. 1993, \apj,
409, 75

\bibitem[Peng, Ford \& Freeman(2004)]{pen04} Peng, E., Ford, H.C., \&
Freeman, K. C. 2004, \apj, 602, 685

\bibitem[Romanowsky \etal (2003)]{rom03} Romanowsky, A.J., \etal\ 2003,
Science, 301, 1696

\bibitem[Ryan-Weber \etal (2004)]{ryweb04} Ryan-Weber, E.V., \etal\ 2004,
\aj, 127, 1431

\bibitem[Saglia \etal (2000)]{sag00} Saglia, R.P., Kronawitter, A., 
Gerhard, O.E., \&  Bender, R., 2000,  \aj, {119}, {153}

\bibitem[Sansom, Hibbard \& Schweizer(2000)]{san00} Sansom, A.E., Hibbard,
J.E., \& Schweizer, F. 2000, \aj, 120, 1946

\bibitem[Sarazin, Irwin \& Bregman (2000)]{sar00} Sarazin, C.L., Irwin, J.
A., \& Bregman J.N. 2000, \apj, 544, L101

\bibitem[Scorza \& Bender(1995)]{sco95} Scorza, C., \& Bender, R. 1995, \aap, 
293, 20

\bibitem[Tonry \etal\ (2001)]{tonry} Tonry, J.L., \etal\ 2001, \apj, 546, 681

\bibitem[Trager \etal (2000)]{tra00} Trager, S.C., Faber, S.M., Worthey, G., 
\& Gonz\'alez, J.J. 2000, \aj, 119, 1645

\bibitem[van Dokkum (2005)]{vdokk05} van Dokkum, P.G. 2005, \aj,
submitted, astro-ph/0506661

\bibitem[Zezas \etal (2003)]{zez03} Zezas, A., Hernquist, L., Fabbiano, G.,
\& Miller, J. 2003, \apjl, 599, L73

\end{thebibliography}
\end{document}